\pdfoutput=1
\documentclass[aip,jcp,reprint,amsmath,amssymb,floatfix]{revtex4-2}
\usepackage[utf8]{inputenc}
\usepackage[T1]{fontenc}
\usepackage{lmodern,microtype}
\usepackage{graphicx,xcolor}
\usepackage{booktabs,array,calc}
\IfFileExists{multirow.sty}{\usepackage{multirow}}{}
\providecommand{\multirow}[3]{#3}
\usepackage[hidelinks]{hyperref}
\providecommand{\tightlist}{\setlength{\itemsep}{0pt}\setlength{\parskip}{0pt}}

\providecommand{\lt}{<}  
\newsavebox{\fitbox}                               
\newcommand{\fitwidth}[1]{\sbox{\fitbox}{#1}\ifdim\wd\fitbox>\linewidth\resizebox{\linewidth}{!}{\usebox{\fitbox}}\else\usebox{\fitbox}\fi}
\DeclareUnicodeCharacter{00B1}{\ensuremath{\pm}}
\DeclareUnicodeCharacter{00C5}{\AA{}}
\DeclareUnicodeCharacter{00D7}{\ensuremath{\times}}
\DeclareUnicodeCharacter{00F3}{\'o}
\DeclareUnicodeCharacter{03B2}{\ensuremath{\beta}}
\DeclareUnicodeCharacter{03B7}{\ensuremath{\eta}}
\DeclareUnicodeCharacter{2192}{\ensuremath{\to}}
\DeclareUnicodeCharacter{2212}{\ensuremath{-}}
\DeclareUnicodeCharacter{221A}{\ensuremath{\surd}}
\DeclareUnicodeCharacter{2264}{\ensuremath{\leq}}
\begin{document}
\title{Jackknife-conformal uncertainty for linear ACE potentials: calibrated forces and silicon properties compared with Bayesian and POPS baselines}
\author{James R. Kermode}
\affiliation{Warwick Centre for Predictive Modelling, School of Engineering, University of Warwick, Coventry CV4 7AL, United Kingdom}
\begin{abstract}
Machine-learned interatomic potentials are misspecified, so Bayesian posterior variances understate their errors. For linear
atomic cluster expansion (ACE) potentials we derive calibrated uncertainties from one misspecification-robust parameter
covariance, the delete-one-configuration jackknife. Scaled by one conformal quantile from a held-out part of the training
set, it gives per-atom force uncertainties close to their target coverage on held-out silicon data and on cracks left out
of the fit; screw dislocations left out are under-covered by the smaller of the two models. With one scale fitted to the forces of a held-out test split,
the same parameter distribution is propagated to 21 properties of silicon, from elastic constants to defect and surface energies,
by implicit differentiation through each relaxation and by committees, and is tested against density functional theory (DFT) without tuning on any property.
For a smaller and a larger model the intervals cover most properties, and the larger model\textquotesingle s are about 4 times
tighter. The Bayesian posterior with one calibrated scale has the wrong shape: it covers fewer properties and has far lower joint
evidence. Pointwise optimal parameter sets (POPS), once their hypercube is numerically defined, performs comparably on the properties, but its per-atom force
regions are not calibrated to a target (0.95 and 0.84 at a nominal 0.90). Where the jackknife fails is relaxed point defects, whose training cells are all hot snapshots: a bias that every configuration
shares is invisible to a jackknife, but it can be measured from the training labels, and as a discrepancy term, devised
after the property results were seen, it brings the coverage close to nominal.

\protect\phantomsection\label{intro}
\end{abstract}
\maketitle
\section{1 Introduction}\label{introduction}

Machine-learned interatomic potentials are used for forces, in molecular dynamics and in active learning where force errors
must be controlled during a simulation, and for the derived properties that a potential is built to predict. Both uses need an
uncertainty. The training labels are deterministic density functional theory (DFT) energies, forces and stresses, so the error of a potential is approximation
error: the model is misspecified. A Bayesian posterior is then far too narrow. Under misspecification it concentrates at the
best-fitting parameters at a rate set by the amount of data, not by the size of the model error, and it collapses to a point as
\(N_{\text{data}}\to\infty\) while the prediction error stays finite\cite{kleijn2012,muller2013,swinburne2025}.

The most widely used alternatives for forces are committees and ensembles of models\cite{musil2019,imbalzano2021,kellner2024} and approximate posteriors on the last layer of a network\cite{bigi2024}. They need several fits, or a
calibration of their own. Conformal prediction gives intervals with guaranteed coverage under exchangeability\cite{vovk2005,tibshirani2019,barber2021}, and Best et al.\cite{best2024} combined it with the Bayesian posterior of linear atomic cluster expansion (ACE) potentials, and committees drawn from it, for
silicon. A single conformal scale covers on average, however, while force errors vary strongly between local environments and are
correlated within a configuration.

In this paper we construct calibrated per-atom force uncertainties for linear ACE potentials\cite{drautz2019} from two parts. The shape is a cluster-robust ("sandwich") covariance of the coefficients\cite{huber1967,white1982}, with each
training configuration as one cluster of correlated rows. We use the variant proposed by Bell and McCaffrey\cite{bell2002} and named CR3 by
Pustejovsky and Tipton\cite{pustejovsky2018} (CV3 in MacKinnon, Nielsen and Webb\cite{mackinnon2023}), the cluster form of the heteroskedasticity-consistent HC3
estimator\cite{mackinnon1985}. CR3 is the delete-one-cluster jackknife: it measures how far the fitted coefficients move when
each configuration is left out. It is computed exactly, without refitting, from the PRESS (predicted residual error sum of squares)
residuals. These are the residuals each row would have if its cluster were left out of the fit, and they follow from the ordinary
residuals by a leverage correction: for one row, division by one minus its leverage; for a cluster, the matrix analogue\cite{allen1974,cook1982}. The per-atom 3×3 block of the covariance gives anisotropic force regions.

The scale comes from split conformal prediction\cite{vovk2005}, which turns the scores of held-out
calibration points into a quantile that covers a new point with a guaranteed probability. One quantile serves every atom: the
jackknife shape already carries the dependence of the error on the environment. A scale learned from the site descriptor
(normalised conformal prediction\cite{papadopoulos2008,lei2018,ho2026}) makes the coverage more even across environments
represented in the training data, but not on environments left out of it, so we report it as a refinement,
not as the method. The scores come from a stratified hold-out of the training set, and a fitted transfer exponent carries them to
the fit on all the data. The open-source \texttt{ace-jax} package\cite{acejax} provides the shape, the hold-out scores and the transfer;
Sections 2.5 and 3.2 describe and test the scale.

Per-atom intervals cannot be combined into an interval on a derived property, a quantity of interest (QoI), because a QoI depends
on every atom and on relaxed geometries. A QoI\textquotesingle s uncertainty has two parts\cite{simmons_inprep}. The
aleatoric part is the sampling error of the simulation that computes the QoI, assessed with statistical convergence tests\cite{grossfield2009}\cite[Sec.~2.3]{lelievre2010}. The epistemic part comes from the uncertainty in the
fit. It is most widely estimated with a committee of potentials, an idea long established in statistics\cite{hoeting1999}. The QoIs here come from relaxations and static calculations, so they have no aleatoric part and their uncertainty is
purely epistemic. For a linear model the consistent object is then a distribution over the coefficients \(\theta\): one draw
gives energies, forces and stresses together, and every QoI follows from it.

We propagate the same CR3 distribution, with one scale fitted to held-out forces, to 21 QoIs of silicon, for potentials fitted
to the GAP-18 database\cite{bartok2018}, the training set of the GAP-18 Gaussian approximation potential. The propagation uses implicit differentiation through every relaxation, including that
of the lattice constant, and is cross-checked by committees run through the unmodified benchmark protocol. We test the intervals
against DFT, one QoI at a time and jointly, and compare them with two baselines: the posterior with one conformal scale, and
pointwise optimal parameter sets (POPS)\cite{swinburne2025}.

Misspecification-aware parameter uncertainty has been propagated to material properties before. Perez et al.\cite{perez2025} propagate
POPS hypercube ensembles of a quadratic spectral neighbour analysis potential (SNAP) for tungsten to phase, elastic, defect and surface properties, re-relaxing
each member, and use the implicit derivative of a relaxed minimum with respect to the parameters\cite{maliyov2025} to predict volumes and a vacancy-migration barrier; the DFT values lie inside the ensemble\textquotesingle s envelope without a calibration
step. We differ in three ways: the shape is the CR3 covariance with one scale calibrated on held-out data, so its size does not
depend on which training rows define a box; the intervals are tested against DFT one QoI at a time and jointly, by coverage, the continuous ranked probability score (CRPS)
and evidence; and the committees run through an unmodified benchmark protocol, beside implicit differentiation of every
relaxation.

We also show where calibrated parameter-space uncertainty fails, and why. A bias that all training configurations of one kind
share is invisible to any delete-one jackknife; in the GAP-18 database\cite{bartok2018} every defect cell is a hot molecular-dynamics snapshot. We
diagnose this bias, measure it from the training labels alone, and add it to the defect QoIs as a discrepancy term, which
leaves the fit unchanged.

Section 2 sets out the methodology: the linear ACE fit (2.1--2.2), automatic relevance determination as a check on the prior (2.3), the CR3 shape (2.4), the per-atom force calibration
(2.5), the baselines (2.6) and the propagation to QoIs (2.7). Section 3 gives the validation statistics and the results: the per-atom forces first, then the global scale that the
propagation uses, then the QoIs. Section 4 discusses them and Section 5 concludes.

\protect\phantomsection\label{theory-model}
\section{2 Methodology}\label{methodology}

\subsection{2.1 Linear ACE: model and observations}\label{linear-ace-model-and-observations}

Each atom \(i\), at position \(r_i\) with species \(Z_i\) and environment \(x_i\) (its neighbours within the cutoff), has a site descriptor \(\phi(x_i)\in\mathbb{R}^p\): the atomic cluster expansion (ACE) product basis\cite{drautz2019}, in
species-major blocks, then pair blocks. The basis, its radial functions and the smoothness prior below are those of
ACEpotentials\cite{witt2023}. The energy is linear in the coefficients \(\theta\),
\begin{equation} E(\{r_i\};\theta)=\sum_i \phi(x_i)\cdot\theta+\sum_i E^0_{Z_i}, \end{equation}
and so is every observation of a configuration:
\begin{gather}b_E=\sum_i\phi(x_i),\\
B_{F,i\alpha}=-\frac{\partial b_E}{\partial r_{i\alpha}},\\
B_{V,\alpha\beta}=-\frac{\partial b_E}{\partial\varepsilon_{\alpha\beta}}\Big|_{\varepsilon=0},
\end{gather}
with \(E^0_Z\) a per-species reference energy, \(r_{i\alpha}\) the Cartesian components of \(r_i\), and \(\varepsilon\) a symmetric strain
applied to positions and cell. The predictions are \(b_E\cdot\theta\), \(B_F\theta\) and the virial \(W=B_V\theta\) (in eV; the stress is
\(-W/\Omega\), with \(\Omega\) the cell volume). Each row of \(b_E\), \(B_F\) or \(B_V\) is called an observable row \(b\).
Each row has a weight \(w\), which combines a factor \(1/\sqrt N\) on the energy and virial rows of an \(N\)-atom configuration, a weight per quantity
(energy, force, virial) and a weight per configuration type, and all rows share one noise scale \(\sigma\). The weights set the
balance between energies, forces and virials, and \(\sigma\) sets the overall noise level. For an energy row the whitened row and
target are \(\psi=b_E\,w/\sigma\) and \(\tilde y=y\,w/\sigma\), with \(y=E_{\text{DFT}}-\sum_iE^0_{Z_i}\) the DFT energy minus the
reference energies. Force and virial rows enter in the same way: each of the \(3N\) rows of \(B_F\) and each of the six independent
(Voigt) rows of \(B_V\) is one row \(b\), with \(\psi=b\,w/\sigma\) and \(\tilde y=y\,w/\sigma\), and its target \(y\) is the matching DFT
force or virial component (\(E^0\) contributes to neither). Stacked over all configurations, the observable rows \(b\) form the design matrix \(\Phi\), and the
\(\psi\) its whitened form \(\Psi\). With least-squares \(E^0\), one column per species carries the atom count on the
energy rows (a fixed, wide prior).

\begin{figure*}[tbp]
\centering
\includegraphics[width=\textwidth,height=0.6\textheight,keepaspectratio]{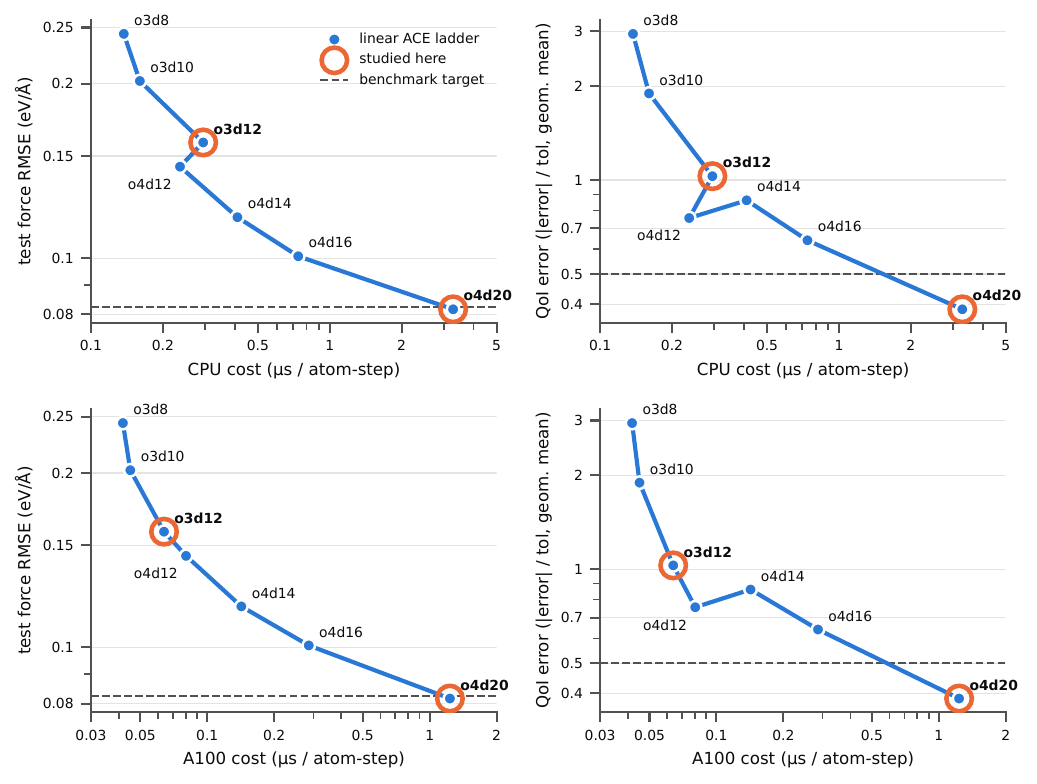}
\caption{The linear ACE ladder of the \protect\texttt{ace-jax} silicon accuracy--cost benchmark: test force RMSE (left) and QoI error (right; the geometric
mean over the silicon QoIs of \textbar error\textbar/tolerance against CASTEP) against cost in LAMMPS, float64: on a CPU node with the
\protect\texttt{ace-jax} C++ kernel (top) and on an NVIDIA A100 GPU with the \protect\texttt{ace-jax} CUDA kernel (bottom; hardware in Section 6). Dashed: the benchmark\textquotesingle s targets, GAP-18\textquotesingle s force RMSE on this split and a QoI error of 0.5. Circled: the two models
studied here.}\label{fig:1}
\end{figure*}

\textbf{The two models.} We study two members of the linear ACE ladder of the \texttt{ace-jax} silicon accuracy--cost benchmark,
fitted to the GAP-18 database\cite{bartok2018} (Figure~\ref{fig:1}). A model oNdM has correlation order N, the largest number of
neighbour factors in a basis function (body order N + 1, the index used for ARD in Section 2.3), and maximum total polynomial
degree M: each one-particle factor t with radial index \(n_t\) and angular degree \(l_t\) contributes \(n_t+\tfrac32 l_t\),
and the sum over the factors of a basis function is at most M. This is the total-degree basis of ACEpotentials, described in
detail by Witt et al.\cite{witt2023}.
Table~\ref{tab:1} lists the silicon QoIs, their DFT (CASTEP) values and the benchmark\textquotesingle s tolerances: 10 \% of the DFT value for the elastic
constants, the bulk modulus, the unstable stacking-fault energies and the decohesion stress; 5 \% for the (100) surface energy;
10 meV/atom for the phase energies; 0.1 eV for the point-defect formation energies; and 0.2 THz for the root-mean-square (RMS) error of the phonon frequencies. The QoI error
of a model is the geometric mean of \textbar error\textbar/tolerance over these QoIs. The ratios span orders of magnitude, so an arithmetic mean
would be set by the one or two worst QoIs; the geometric mean, the exponential of the mean log ratio, counts a factor-of-two
improvement equally on every QoI. o3d12 (223 coefficients) is the cheaper model, with a
test force RMS error (RMSE) of 0.158 eV/Å and a typical QoI error of one tolerance (1.03). o4d20 (5476 coefficients)
is the only size of the ladder that meets both of the benchmark\textquotesingle s targets, a test force RMSE no larger than GAP-18\textquotesingle s 0.0825 eV/Å
on this split (0.0816 eV/Å) and a QoI error below 0.5 (0.39); o3d12 is 11× cheaper on a
CPU node and 19× cheaper on an NVIDIA A100 GPU. The pair spans the useful range of the ladder: a cheap model
whose intervals must be wide to be honest, and an accurate one whose intervals should be narrow. The two columns of Figure~\ref{fig:1} also rank the sizes differently: o4d12 has a larger force error than o4d14 but a smaller QoI
error (0.76 against 0.86). Force accuracy alone does not predict QoI accuracy, which is why the
uncertainty is propagated to the QoIs themselves.

\begin{table*}[htbp]
\centering\small\setlength{\tabcolsep}{3pt}
\caption{The silicon QoIs: DFT (CASTEP) reference value, unit, benchmark tolerance, and the value, error (model − DFT) and relative error (error / DFT) of the two ACE models. A tolerance in \% is relative to the DFT value. The 10 meV/atom phase-energy tolerance is most of the small hexagonal-diamond value; the relative error is the more telling measure there. The phonon entry is the RMS difference of the phonon frequencies from DFT, so it has no DFT value or error and is not scored for coverage.}\label{tab:1}
\fitwidth{\begin{tabular}{@{}llllllllll@{}}
\toprule
\multirow{2}{*}{QoI} & \multirow{2}{*}{DFT} & \multirow{2}{*}{unit} & \multirow{2}{*}{tolerance} & \multicolumn{3}{l}{%
ACE o3d12} & \multicolumn{3}{l@{}}{%
ACE o4d20} \\
& & & & value & error & error (\%) & value & error & error (\%) \\
\midrule
\(C_{11}\) & 153.3 & GPa & 10 \% & 130.3 & −23.0 & −15.0 & 151.3 & −2.0 & −1.3 \\
\(C_{12}\) & 56.3 & GPa & 10 \% & 60.7 & +4.5 & +8.0 & 55.2 & −1.1 & −1.9 \\
\(C_{44}\) & 72.2 & GPa & 10 \% & 73.4 & +1.2 & +1.7 & 70.6 & −1.6 & −2.2 \\
bulk modulus \(B\) & 88.6 & GPa & 10 \% & 83.9 & −4.7 & −5.3 & 87.2 & −1.4 & −1.6 \\
E(hexagonal diamond) − E(diamond) & 11.5 & meV/atom & 10 meV/atom & 13.0 & +1.6 & +13.5 & 11.1 & −0.4 & −3.1 \\
E(bc8) − E(diamond) & 163.0 & meV/atom & 10 meV/atom & 161.7 & −1.3 & −0.8 & 159.9 & −3.1 & −1.9 \\
E(st12) − E(diamond) & 172.5 & meV/atom & 10 meV/atom & 155.4 & −17.0 & −9.9 & 168.5 & −4.0 & −2.3 \\
E(β-Sn) − E(diamond) & 321.7 & meV/atom & 10 meV/atom & 319.6 & −2.1 & −0.6 & 320.6 & −1.1 & −0.3 \\
E(simple hexagonal) − E(diamond) & 344.1 & meV/atom & 10 meV/atom & 340.0 & −4.1 & −1.2 & 341.6 & −2.5 & −0.7 \\
E(hcp) − E(diamond) & 561.4 & meV/atom & 10 meV/atom & 562.7 & +1.4 & +0.2 & 560.3 & −1.0 & −0.2 \\
E(bcc) − E(diamond) & 572.2 & meV/atom & 10 meV/atom & 541.4 & −30.8 & −5.4 & 573.1 & +0.9 & +0.2 \\
E(fcc) − E(diamond) & 589.7 & meV/atom & 10 meV/atom & 569.9 & −19.7 & −3.3 & 592.9 & +3.3 & +0.6 \\
vacancy formation & 3.673 & eV & 0.1 eV & 2.792 & −0.881 & −24.0 & 3.562 & −0.111 & −3.0 \\
tetrahedral interstitial formation & 3.910 & eV & 0.1 eV & 2.172 & −1.738 & −44.5 & 3.311 & −0.599 & −15.3 \\
hexagonal interstitial formation & 3.724 & eV & 0.1 eV & 2.277 & −1.448 & −38.9 & 3.312 & −0.412 & −11.1 \\
dumbbell interstitial formation & 3.662 & eV & 0.1 eV & 2.481 & −1.181 & −32.2 & 3.290 & −0.372 & −10.1 \\
(100) surface energy, relaxed & 2.173 & J/m\textsuperscript{2} & 5 \% & 2.108 & −0.064 & −3.0 & 2.133 & −0.040 & −1.8 \\
unstable stacking fault, glide & 1.742 & J/m\textsuperscript{2} & 10 \% & 1.819 & +0.077 & +4.4 & 1.870 & +0.128 & +7.3 \\
unstable stacking fault, shuffle & 1.608 & J/m\textsuperscript{2} & 10 \% & 1.021 & −0.587 & −36.5 & 1.461 & −0.147 & −9.2 \\
maximum (100) decohesion stress, unrelaxed & 0.189 & eV/Å\textsuperscript{3} & 10 \% & 0.200 & +0.011 & +5.7 & 0.184 & −0.005 & −2.6 \\
phonon frequencies, RMS against DFT & -- & THz & 0.2 THz & 0.286 & -- & -- & 0.153 & -- & -- \\
\bottomrule
\end{tabular}}
\end{table*}

\subsection{2.2 Prior, posterior and evidence}\label{prior-posterior-and-evidence}

The prior is \(\theta\sim\mathcal N(0,\Lambda^{-1})\) with the smoothness prior
\begin{gather}\Lambda=\frac{D^2}{\sigma_c^2},\\
D=\operatorname{diag}(\Gamma_j),\\
\Gamma_j=\sum_t\Big[\Big(\frac{l_t}{w_l}\Big)^4+\Big(\frac{n_t}{w_n}\Big)^4\Big],
\end{gather}
where \(\sigma_c\) is the prior scale of the coefficients and the sum runs over the one-particle factors \(t\) of basis function \(j\), with radial
index \(n_t\) and angular degree \(l_t\) (the algebraic prior of ACEpotentials\cite{witt2023}, \(w_l=2/3\), \(w_n=1\)), so rough basis
functions are damped. The smoothness weights \(D\) do not depend on any hyperparameter; Sections 2.3 and 2.4 use them to scale
the coefficients.
This is Bayesian linear regression (BLR). The posterior is Gaussian, with mean \(\hat\theta\), the maximum a posteriori
(MAP) estimate,
\begin{gather}A=\sum_{\text{rows}}\psi\psi^{\mathsf T}+\Lambda,\\
\hat\theta=A^{-1}\sum_{\text{rows}}\psi\tilde y,\\
\operatorname{Cov}(\theta\mid\mathcal D)=A^{-1},
\end{gather}
and the hyperparameters \(h=(\sigma_c,\sigma)\) maximise the evidence (for o3d12 the maximum lies on the upper bound
\(\sigma_c=10^4\) of the search, and the fit is used there; Section 3.7)
\begin{multline}\log p(\mathcal D\mid h)=-\frac{\lVert y^w\rVert^2}{2\sigma^2}+\tfrac12 c^{\mathsf T}A^{-1}c-\tfrac12\log|A|\\
+\tfrac12\log|\Lambda|-n\log\sigma ,
\end{multline}
up to a constant, with \(\mathcal D\) the training data, \(y^w\) the weighted targets (\(y\,w\)), \(n\) the number of rows,
and \(c=\sum_{\text{rows}}\psi\tilde y=A\hat\theta\).

Near-interpolating fits leave the condition number \(\operatorname{cond}(A)\) beyond \(1/\epsilon_{\text{mach}}\) (\(\epsilon_{\text{mach}}\)
the float64 unit roundoff), so the evidence and the posterior are computed by QR factorisation. \(R_w\) is the triangular factor of the weighted
design (\(\Phi^w=Q_wR_w\), rows \(b\,w\)), accumulated by streamed updating QRs, and
\begin{gather}\begin{bmatrix}\Lambda^{1/2}\\ R_w/\sigma\end{bmatrix}=Q_AR_A,\\
R_A^{\mathsf T}R_A=A,
\end{gather}
which works at \(\operatorname{cond}(\Phi)\) instead of \(\operatorname{cond}(\Phi)^2\).

\protect\phantomsection\label{theory-ard}
\subsection{2.3 Sensitivity to the prior: automatic relevance determination}\label{sensitivity-to-the-prior-automatic-relevance-determination}

Automatic relevance determination (ARD) replaces the single \(\sigma_c\) by one log-precision per body order,
\begin{equation} \Lambda=D\,\operatorname{diag}\big(e^{a_{k(j)}}\big)\,D, \end{equation}
with \(k(j)\) the body order of basis function \(j\) and \(D\) the smoothness weights of Section 2.2. The BLR fit is ARD at
\(a_k=-2\log\sigma_c\) for every \(k\). The \(a_k\) are fitted by type-II maximum likelihood in the scaled system
\(S=D^{-1}AD^{-1}\), whose prior part is \(\operatorname{diag}(e^{a_{k(j)}})\). The scaling is by \(D\), not by \(\Lambda^{1/2}\),
because \(D\) does not change with the \(a_k\): the data part of \(S\) is computed once, and only its diagonal changes during the
maximisation. In sequential mode, used here, the noise scale \(\sigma\) stays at the BLR fit\textquotesingle s value and only the
\(a_k\) are fitted.

\textbf{QR evidence.} As for the BLR fit (Section 2.2), the evidence, its gradient and Hessian, and the posterior are computed
from the QR factor
\begin{gather}\begin{bmatrix}\operatorname{diag}\big(e^{a_{k(j)}/2}\big)\\ R_wD^{-1}/\sigma\end{bmatrix}=Q_SR_S,\\
R_S^{\mathsf T}R_S=S,
\end{gather}
whose condition number is \(\sqrt{\operatorname{cond}(S)}\). The posterior stores \(R_S^{\mathsf T}\) as its Cholesky factor, so the CR3 jackknife and the per-atom force variances are unchanged. The \(a_k\) are bounded below at
\(a_{\text{floor}}=\log(\lambda_{\max}/10^{22})\), with \(\lambda_{\max}\) the largest eigenvalue of the data part of \(S\)
(\(D^{-1}\Psi^{\mathsf T}\Psi D^{-1}\)) at the starting noise scale, and an endpoint with every \(a_k\) on the floor and the gradient pointing below it
is flagged as degenerate and refused.

\begin{table*}[htbp]
\centering\small\setlength{\tabcolsep}{3pt}
\caption{ARD against the benchmark BLR fits. ARD log-precision per body order at the endpoint
(* on the floor), the floor, and the single log-precision of the BLR prior (−2 log \(\sigma_c\); * on the upper bound \(\sigma_c=10^{4}\) of
the evidence search); test-split RMSE (E and V per atom) over the rows the calibration keeps, so without the force components
that vanish by site symmetry, which is why the force RMSE is slightly above the benchmark\textquotesingle s value in Section 2.1;
test-split \(\lambda_F\); and the propagated CR3 × \(\lambda_F\) coverage of the 20 scored QoIs.}\label{tab:2}
\fitwidth{\begin{tabular}{@{}lllllllllllllllll@{}}
\toprule
\multirow{2}{*}{model} & \multicolumn{4}{l}{%
ARD log-precision \(a_k\) (body order \(k\))} & \multirow{2}{*}{ARD floor} & \multirow{2}{*}{BLR \(a\)} & \multicolumn{2}{l}{%
E RMSE (meV/atom)} & \multicolumn{2}{l}{%
F RMSE (eV/Å)} & \multicolumn{2}{l}{%
V RMSE (meV/atom)} & \multicolumn{2}{l}{%
test-split \(\lambda_F\)} & \multicolumn{2}{l@{}}{%
coverage, 95 \% (68 \%)} \\
& 2 & 3 & 4 & 5 & & & ARD & BLR & ARD & BLR & ARD & BLR & ARD & BLR & ARD & BLR \\
\midrule
o3d12 & -24.5* & -20.6 & -18.7 & -- & -24.5 & -18.4* & 15.89 & 15.90 & 0.1588 & 0.1589 & 112.9 & 113.2 & 5.88 & 5.82 & 17 (14) & 17 (14) \\
o4d20 & -12.0 & -12.3 & -8.7 & -6.4 & -14.1 & -7.8 & 3.38 & 3.50 & 0.0819 & 0.0819 & 32.4 & 32.6 & 6.18 & 6.17 & 15 (12) & 15 (12) \\
\bottomrule
\end{tabular}}
\end{table*}

Neither fit is degenerate. For o3d12 only the two-body precision sits on the floor, and for o4d20 every \(a_k\) is
interior. ARD weakens the prior relative to the BLR\textquotesingle s single \(\sigma_c\) on most body orders (on all of them for o3d12; for
o4d20 only the 5-body block is held more tightly), yet the means match the BLR fits on the held-out test split, and the test-split
scales and the propagated coverage agree. Every result in this paper, per-atom and QoI, therefore uses one posterior per model:
the BLR fit, which is the benchmark model. ARD is reported only as this check. The per-atom calibration of Section 2.5 run on
the ARD fits covers 0.87 (o3d12) and 0.90 (o4d20) of the test atoms at a nominal 0.90, against
0.87 and 0.90 on the BLR fits (Section 3.2).

\protect\phantomsection\label{theory-cr3}
\subsection{2.4 Misspecification-robust shape: the CR3 jackknife}\label{misspecification-robust-shape-the-cr3-jackknife}

\textbf{POPS.} The closest existing approach for linear potentials is POPS\cite{swinburne2025}. For each training row
\(i\), with residual \(r_i\) and leverage \(h_i=\psi_i^{\mathsf T}A^{-1}\psi_i\), the pointwise optimal parameter set is
\(\hat\theta+\delta_i\), with \(\delta_i=A^{-1}\psi_i\,r_i/h_i\) the smallest change (in the \(A\) metric) that makes the model
reproduce that row exactly. The spread of the \(\delta_i\), as an ensemble or as a hypercube in their principal directions, is
the uncertainty. It does not shrink as data are added, because each \(\delta_i\) stays finite while the model is misspecified.

\textbf{Comparison.} CR3 below is built from the same per-row influence, \(A^{-1}\psi_i r_i\), but asks the opposite question:
how far the fit moves when a configuration is removed, not how far it must move to fit a row exactly. For one row the two differ
only in the leverage factor, \(1/(1-h_i)\) for deletion against \(1/h_i\) for an exact fit. CR3 is therefore largest for rows the
fit leans on, and POPS for rows the fit hardly sees, whose corrections are unbounded as \(h_i\to0\) (Section 3.7). CR3 also deletes
whole configurations, so that the correlated energy, force and virial rows of one configuration move together, and it is a
covariance that is calibrated by one scale on held-out data (Section 3.3); POPS has no calibration step. Like any sampling variance,
the CR3 covariance shrinks as data are added, which is why the scale is needed: CR3 supplies the shape, and the held-out data the
size.

\begin{figure*}[tbp]
\centering
\includegraphics[width=\textwidth,height=0.6\textheight,keepaspectratio]{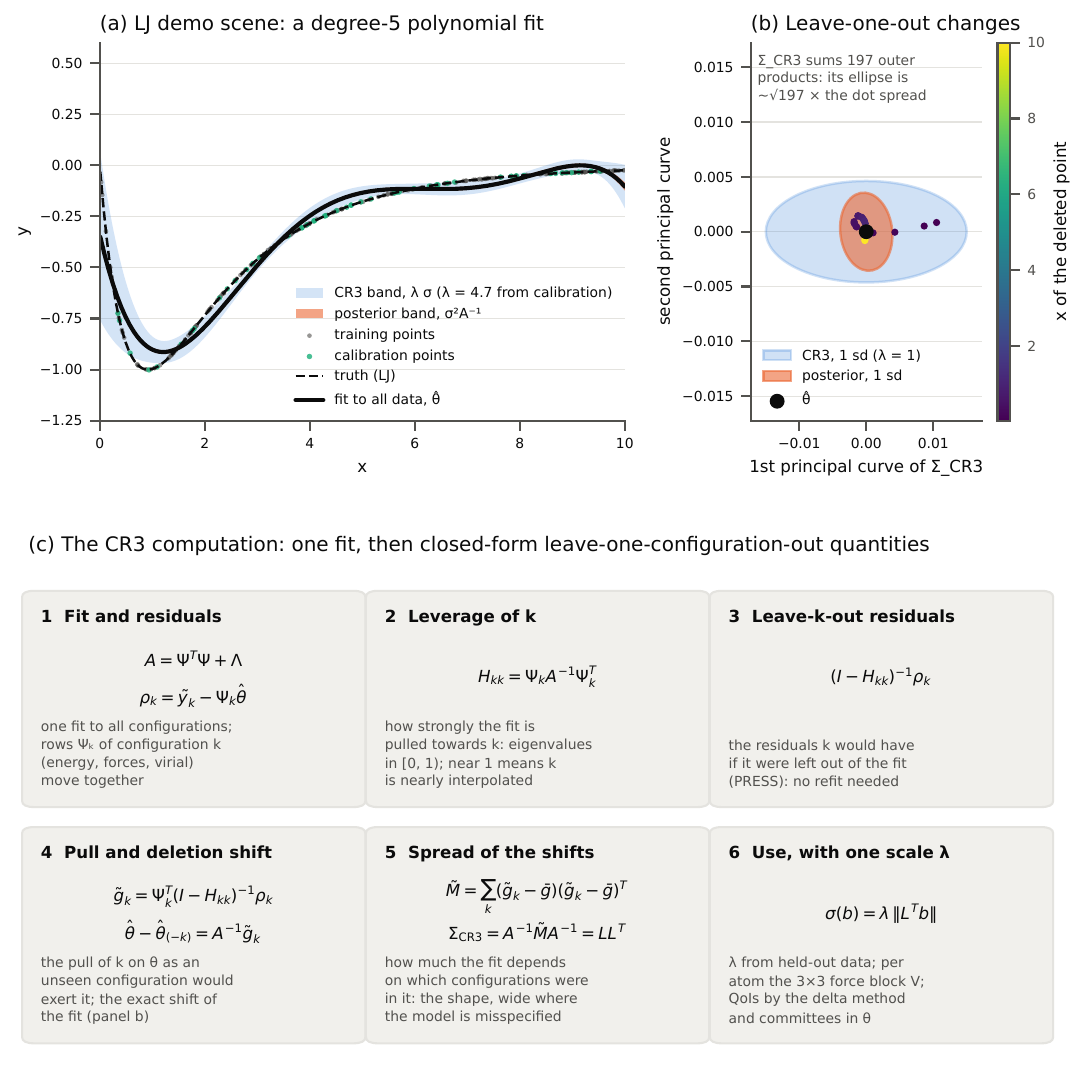}
\caption{The CR3 construction on the Lennard-Jones (LJ) example of the interactive regression demo\cite[scene lj-cr3]{demos}, computed exactly. (a) A degree-5 polynomial, fitted by Bayesian ridge regression with evidence-optimised
hyperparameters to 197 nearly noiseless samples (noise standard deviation 0.001, as for DFT labels) of the smooth part of a
Lennard-Jones curve, cannot follow the steep repulsive side and the well: the error is approximation error, not noise. The
posterior band (orange; \(\sigma^2A^{-1}\) in the demo\textquotesingle s unwhitened rows) is barely visible. The CR3 band (blue), scaled by \(\lambda\) from 50 held-out calibration
points, is wide where the fit is wrong: on the steep side near \(x=0\) and at the right-hand end. (b) Each dot is the change of
the fitted curve when one training point is left out, \(-A^{-1}\tilde g_k\), in the two leading principal curves of the CR3
covariance of the prediction; the closed form reproduces the refit to the precision the basis allows. The points on the steep
side and in the well (\(x<1\)) move the fit most. The CR3 covariance is the sum of these shifts\textquotesingle{} outer products over the 197
training points, so its ellipse is about √197 times the spread of single dots; the posterior ellipse is smaller. (c) The
computation for the real models, with the reading of each quantity: one fit, then per configuration its leverage, its
leave-out residuals and its pull, and finally the spread of the pulls.}\label{fig:2}
\end{figure*}

\textbf{CR3.} For a cluster \(k=1,\dots,n_{\text{cl}}\) of rows (here: one whole configuration with its energy, force and virial rows) with
whitened rows \(\Psi_k\) (the cluster\textquotesingle s rows \(\psi^{\mathsf T}\) stacked), whitened targets \(\tilde y_k\) and residuals \(\rho_k=\tilde y_k-\Psi_k\hat\theta\), the cluster hat block and the leverage-corrected (PRESS)
score are
\begin{gather}H_{kk}=\Psi_kA^{-1}\Psi_k^{\mathsf T}=W_k^{\mathsf T}W_k,\\
W_k=R_S^{-\mathsf T}D^{-1}\Psi_k^{\mathsf T},\\
\tilde g_k=\Psi_k^{\mathsf T}(I-H_{kk})^{-1}\rho_k .
\end{gather}
Each object has a direct reading.

\begin{itemize}
\tightlist
\item
  \(H_{kk}\) is the cluster\textquotesingle s block of the hat matrix: it maps the cluster\textquotesingle s targets to its own fitted values, so it measures how
  far the fit is pulled towards that configuration. Because \(\Lambda\succ0\), its eigenvalues lie in \([0,1)\). An eigenvalue near 0
  means the configuration is predicted by the rest of the data; one near 1 means the fit nearly interpolates it along that direction,
  and its residual there underestimates the error it would have if it were left out.
\item
  \(W_k\) is the cluster\textquotesingle s rows expressed in whitened coefficient coordinates, scaled by the smoothness weights (\(D\)) and by the Cholesky
  factor of \(S\) (Section 2.3). In these coordinates the posterior is isotropic, so \(H_{kk}=W_k^{\mathsf T}W_k\) is a plain Gram
  matrix; computing \(W_k\) costs one triangular solve per cluster.
\item
  \(\tilde g_k\) is the cluster\textquotesingle s contribution to the gradient of the least-squares objective, \(\Psi_k^{\mathsf T}\rho_k\),
  evaluated with its leave-cluster-out (PRESS) residuals \((I-H_{kk})^{-1}\rho_k\) instead of its ordinary residuals. It is the pull of
  configuration \(k\) on the coefficients as an unseen configuration would exert it.
\end{itemize}

With the hyperparameters fixed, deleting cluster \(k\) moves the mean by exactly \(\hat\theta-\hat\theta_{(-k)}=A^{-1}\tilde g_k\)
(the cluster DFBETA\cite[Sec.~3.6]{cook1982}). The spread of these shifts over clusters is the CR3 covariance. The
centred meat and the
delete-one-cluster (CR3) covariance are
\begin{gather}\tilde M=\sum_k(\tilde g_k-\bar g)(\tilde g_k-\bar g)^{\mathsf T},\\
\Sigma_{\text{CR3}}=A^{-1}\tilde MA^{-1}=D^{-1}\tilde Q\tilde Q^{\mathsf T}D^{-1},\\
\tilde Q=S^{-1}D^{-1}[\tilde g_1-\bar g,\dots].
\end{gather}
with \(\bar g=n_{\text{cl}}^{-1}\sum_k\tilde g_k\). Centring removes the common offset of the pulls: for the ordinary pulls
\(\Psi_k^{\mathsf T}\rho_k\) it is the prior\textquotesingle s gradient (\(\sum_k\Psi_k^{\mathsf T}\rho_k=\Lambda\hat\theta\) at the MAP); the
leverage-corrected \(\tilde g_k\) are centred on their own mean.
\(\Sigma_{\text{CR3}}\) is a Huber--White sandwich. With one residual variance \(s^2\) for every row, the expected uncorrected meat
is \(s^2(A-\Lambda)\), so the sandwich is \(s^2A^{-1}(A-\Lambda)A^{-1}\), which tends to \(s^2A^{-1}\) for a weak prior: the
homoscedastic special case. It is stored as a factor
\(R\) with \(RR^{\mathsf T}=\tilde Q\tilde Q^{\mathsf T}\) (rank ≤ number of clusters), so that
\begin{gather}\Sigma=LL^{\mathsf T},\\
L=D^{-1}R\in\mathbb R^{p\times r},
\end{gather}
with \(r\) the rank. For any observable row \(b\), \(\sigma(b)=\lambda\,\lVert L^{\mathsf T}b\rVert\), where \(\lambda\) is a scale
calibrated on held-out data (Section 3.3). The CR3 covariance measures how much a
prediction depends on which configurations were in the fit: large where the fit leans on configurations it reproduces
poorly. It is a parameter variance used as a proxy for approximation error, which is why it needs a scale.

\subsection{2.5 Calibrated per-atom forces: the CR3 shape and one conformal quantile}\label{calibrated-per-atom-forces-the-cr3-shape-and-one-conformal-quantile}

This section describes the per-atom force uncertainty introduced in this work. Every predicted force gets
a calibrated uncertainty: a shape times one scale. The shape is the per-atom 3×3 block of the CR3
covariance, \(V(x)_{\alpha\beta}=u_\alpha^{\mathsf T}RR^{\mathsf T}u_\beta\) with \(u_\alpha=D^{-1}B_{F,i\alpha}^{\mathsf T}\) built
from the atom\textquotesingle s three force rows (\(\alpha\in\{x,y,z\}\)), and \(v=\operatorname{tr}V\). The scale is one conformal quantile
of scores on held-out training configurations.

\textbf{Hold-out.} Training configurations are split 80/20 into \(T_{\text{fit}}\) and \(T_{\text{val}}\), stratified so that
rare, strongly distorted or off-coordinated environments appear in both parts (the \texttt{ace-jax} default). The posterior
\(P_{\text{fit}}\) on \(T_{\text{fit}}\) keeps the prior and noise scale of the full BLR fit (they are not refitted on
\(T_{\text{fit}}\)), so it differs from the final posterior only in its data. It has its own CR3 shape \(V_{\text{fit}}\), predicts \(T_{\text{val}}\), so both the force
error \(e_i=F_i^{\text{DFT}}-F_i^{\text{fit}}\in\mathbb R^3\) and the shape are out of sample. The anisotropic score is
\begin{gather}s_i=\sqrt{e_i^{\mathsf T}\big(V_{\text{fit}}(x_i)+\epsilon_i I\big)^{-1}e_i},\\
\epsilon_i=10^{-3}\,\tfrac13\operatorname{tr}V_{\text{fit}}(x_i),
\end{gather}
and \(s/\lambda\sim\chi_3\) (the chi distribution with three degrees of freedom) under \(e\sim\mathcal N(0,\lambda^2V)\).

\textbf{Transfer to the full training set.} The final posterior is fitted on all of \(T\), about 25 \% more data than
\(P_{\text{fit}}\). A second hold-out fit \(P_{\text{fit2}}\) on a subset of \(T_{\text{fit}}\) scores the same
\(T_{\text{val}}\) atoms. With \(N\), \(N_{\text{fit}}\), \(N_{\text{fit2}}\) the numbers of
configurations behind the final, first and second fits, and \(\lambda_1,\lambda_2\) the RMS scores of \(P_{\text{fit}}\) and \(P_{\text{fit2}}\),
\(\beta=\operatorname{clip}\big(\log(\lambda_1/\lambda_2)/\log(N_{\text{fit}}/N_{\text{fit2}}),0,\tfrac12\big)\) and every score is
multiplied by \(f_{\text{tr}}=(N/N_{\text{fit}})^\beta\). \(\beta=0\) is the variance-dominated limit and \(\tfrac12\) the bias-dominated one.

\textbf{One posterior.} The final posterior of this stage is the BLR posterior that the QoI propagation uses (Section 2.7): \texttt{ace-jax}
computes the stage at fixed hyperparameters (\texttt{-\/-ard-prior\ fixed}, ARD at the BLR\textquotesingle s single \(a=-2\log\sigma_c\),
Section 2.3) and the QR solver. Against the CR3 posterior built independently at the BLR fit, its mean differs by
\(2\times10^{-6}\) (o3d12) and \(8\times10^{-10}\) (o4d20) in relative norm, and its CR3 covariance by
\(4\times10^{-6}\) and \(3\times10^{-8}\).

\textbf{The scale.} The conformal quantile \(q\) is the \(\lceil(n+1)(1-\alpha)\rceil/n\) empirical quantile of the \(n\) scores
\(s_i\) of the \(T_{\text{val}}\) atoms, at level \(1-\alpha\) (default \(\alpha=0.1\)). It is atom-weighted: a quantile that counts
each configuration once in total is dominated by the many small, well-predicted cells and under-covers (Section 3.2).

\textbf{Per-atom outputs.} The conformal region for a force error \(\Delta F\) is
\(\{\Delta F:\Delta F^{\mathsf T}(V+\epsilon I)^{-1}\Delta F\le q^2\}\), the CR3 ellipsoid scaled to the quantile. Under
exchangeability of the calibration and target configurations it covers a new atom with probability \(\approx1-\alpha\). The
Gaussian reading is the force standard deviation \(\sigma_{F,i}=(q/\chi_3^{-1}(1-\alpha))\sqrt v\), with \(\chi_3^{-1}\) the
\(\chi_3\) quantile function, and covariance \(\sigma_{F,i}^2V/v\).

\textbf{A learned scale, as a refinement.} The CR3 shape is a quadratic form in the atom\textquotesingle s force rows, so it already grows with
the distance of the environment from the training descriptors. In that it is related to the D-optimality extrapolation grade
used to select training data for linear potentials\cite{podryabinkin2017,lysogorskiy2023}, a leverage of the
atom\textquotesingle s descriptor against the training set; the CR3 shape weights that leverage by the residuals the fit leaves, configuration by
configuration, and the quantile calibrates it to a coverage. A scale that depends on the environment should therefore model the
ratio of the error to the shape, the score \(s\), not novelty itself. We test one learned from the site descriptor (normalised
conformal prediction\cite{papadopoulos2008,lei2018}). Let \(\phi_i\) be atom \(i\)\textquotesingle s whitened principal components
of the normalised site descriptor, from a principal component analysis of training atoms only. On one half of the \(T_{\text{val}}\) configurations a ridge
regression \(f(\phi)=\beta_0+\beta^{\mathsf T}\phi+\gamma^{\mathsf T}\phi^{2}\) of \(\log s_i^2\) is fitted (each configuration
weighted once, the ridge by five-fold cross-validation over configurations) and clipped to the range it takes on the
calibration atoms. On the other half the normalised scores \(s_i\,e^{-f(\phi_i)/2}\) give one quantile \(q_f\), and the region is
\(s\le q_f\,e^{f(\phi)/2}\). The split is repeated over five seeds. Ho, Ortner and Wang\cite{ho2026} learn an
environment-dependent scale in the same spirit: a neural network of MACE descriptors that multiplies a last-layer uncertainty,
fitted with a weighted absolute-error loss rather than as a conformal quantile.

\textbf{Why the propagation does not use this scale.} The conformal quantile tempers force \emph{outputs} atom by
atom. It does not define a covariance in \(\theta\), and a QoI depends on every atom and on relaxed geometries. Propagation therefore uses one
global \(\lambda\) (Section 3.3). Section 3.2 tests the per-atom force uncertainties directly.

\subsection{2.6 Baselines}\label{baselines}

We compare the CR3 propagation with three baselines. Each changes one ingredient, the shape of the covariance or the way its
size is set, so that the comparison shows which ingredient matters.

The first baseline, B1, is the approach of Best et al.\cite{best2024}. It keeps the Bayesian posterior as the shape, \(\Sigma=A^{-1}\),
and corrects its size with one conformal quantile \(\hat q\). The quantile is taken from the absolute standardised scores
\(|z|\) of the held-out test split (Section 3.3, here computed with the posterior shape) at the 68.27 \% level, so that
\(\pm\hat q\sigma\) is the \(1\sigma\) interval, pooled over energies, forces and virials, with every configuration given equal weight.
B1 therefore uses the same held-out data as the global scale of our method. It differs mainly in the shape, so it tests whether
the CR3 covariance describes the errors better than the posterior; it also forms its single scale differently (a quantile of
\(|z|\) pooled over E, F and V, against an RMS over force components).

The second baseline, B2, is POPS\cite{swinburne2025}, as implemented in \texttt{ace-jax}\cite{acejax} by a port of the reference
package POPSRegression\cite{popsregression}. For each training row \(i\) the correction is
\(\delta_i=A_\mu^{-1}\psi^w_i\,r_i/h_i\), with leverage \(h_i=\psi_i^{w\mathsf T}A_\mu^{-1}\psi^w_i\). Here \(\psi^w_i=b_i w_i\) carries the
row weights but no noise scale, \(r_i\) is the row\textquotesingle s residual at the ridge solution, and \(A_\mu=M+\mu\operatorname{diag}(\Gamma_j^2)\),
with \(M=\sum_i\psi^w_i\psi_i^{w\mathsf T}\) and the ridge parameter \(\mu\) chosen on a path. Members are drawn uniformly in the
hypercube spanned by the principal directions of the \(\delta_i\), with bounds from their extreme projections, as in
Swinburne and Perez\cite{swinburne2025}. Following the reference package, the box is built only from the rows at or above the median leverage
(\texttt{leverage\_percentile\ =\ 50}, its default; a box from every row is unphysical here, Section 3.7). The ridge is selected
per quantity by the CRPS on a hold-out of the training set, over relative ridges from \(10^{-2}\) down to \(10^{-10}\) of
the largest eigenvalue, a decade above the relative ridge of about \(10^{-11}\) below which the box is set by roundoff
(Section 3.7). We use \texttt{ace-jax}\cite{acejax} version 0.5.3, which includes the
numerical corrections that this study prompted. B2 has no calibration step, so it tests a misspecification-aware shape without a
fitted scale.

The third baseline, B3, is the CR3 covariance with \(\lambda=1\). It isolates the effect of the scale: the difference between B3
and the headline is entirely the calibration of Section 3.3.

Fisher et al.\cite{fisher2026} take the conformal step further. They calibrate joint conformal sets for energies, forces and virials
(Bonferroni-corrected boxes, Mahalanobis ellipsoids, and losses under conformal risk control) and propagate them to derived
quantities such as elastic constants and vacancy formation energies, where modelling the correlations captures error
cancellation. B1 and our CR3 method use one scalar scale instead. We do not use multivariate conformal prediction here; combining
it with the CR3 shape is left to future work.

\protect\phantomsection\label{method-prop}
\subsection{2.7 Propagation to QoIs}\label{propagation-to-qois}

The distribution propagated is \(\theta\sim\mathcal N(\hat\theta,\lambda^2LL^{\mathsf T})\); one sample gives energy, forces and
stress consistently. Per-atom force ellipsoids are never combined.

\subsubsection{2.7.1 The QoIs}\label{the-qois}

The 21 QoIs are those of the silicon test suite of Bartók et al.\cite{bartok2018}, with their DFT (CASTEP) reference values, run with
the tolerances of the \texttt{ace-jax} silicon benchmark (a Python 3 port of that test framework):
\(C_{11}\), \(C_{12}\), \(C_{44}\), \(B\); eight phase energies relative to diamond (compressed hcp excluded as extrapolation);
vacancy and three interstitial formation energies (216-atom cells); the relaxed (100) surface energy; the maximum unrelaxed
decohesion stress; the glide and shuffle unstable stacking-fault energies; the phonon RMS against DFT. Every protocol starts
from the model\textquotesingle s own relaxed diamond cell, so no QoI is linear in \(\theta\). The phonon RMS is a distance to DFT and has no DFT
value: it is left out of coverage (\(n=20\)).

\subsubsection{2.7.2 Delta method by implicit differentiation}\label{delta-method-by-implicit-differentiation}

Implicit differentiation of a relaxed minimum with respect to the parameters of an interatomic potential is the method of
Maliyov, Grigorev and Swinburne\cite{maliyov2025}; here it is applied to every relaxation of the benchmark, the lattice constant included.
Write the energy of a structure with coordinates \(x\) (positions, and cell where relaxed) as \(E=b_E(x)\cdot\theta+E^0(x)\), with
\(b_E(x)\) its energy row (Section 2.1) and \(E^0(x)\) the reference energies. A QoI \(Q(\theta)\) is computed at a relaxed geometry
\(x^*(\theta)\) solving \(\nabla_xE=0\), so
\begin{gather}\frac{dx^*}{d\theta}=-\mathcal H^{-1}\frac{\partial\nabla_xE}{\partial\theta}=-\mathcal H^{-1}\Big(\frac{\partial b_E}{\partial x}\Big)^{\mathsf T},\\
\frac{dQ}{d\theta}=\frac{\partial Q}{\partial\theta}+\frac{\partial Q}{\partial x}\frac{dx^*}{d\theta},
\end{gather}
with \(\mathcal H=\nabla_x^2E\) the Hessian at \(x^*(\hat\theta)\). Then \(\sigma_Q=\lambda\lVert L^{\mathsf T}G\rVert\), with \(G=dQ/d\theta\)
the QoI\textquotesingle s row.

\begin{itemize}
\tightlist
\item
  \textbf{Lattice constant.} The diamond lattice constant \(a_0\) makes the diamond-cell energy \(E_{\text{dia}}(a)\) stationary, so
  \begin{gather}\frac{da_0}{d\theta}=-\frac{1}{E_{aa}}\frac{\partial b_{E,\text{dia}}}{\partial a},\\
E_{aa}=\frac{d^2E_{\text{dia}}}{da^2}\Big|_{a_0}.
\end{gather}
\item
  \textbf{Energies at a minimum} (phase, defect, surface energies). By the envelope theorem the relaxed coordinates add nothing to the derivative of the relaxed
  energy \(E^*\): it is the energy row at \(x^*\) plus the response of the cell the protocol builds from \(a_0\),
  \begin{gather}\frac{dE^*}{d\theta}=b_E(x^*)+\frac{dE}{da_0}\frac{da_0}{d\theta},\\
\frac{dE}{da_0}=\frac{1}{a_0}\,\theta\cdot\frac{\partial b_E}{\partial\ln a}=-\frac{\operatorname{tr}W}{a_0},
\end{gather}
  with \(W\) the virial of the relaxed structure. The relaxed geometries are the harness\textquotesingle s own, recorded for the mean model. For
  example, the vacancy formation energy has the row
  \begin{equation} G=b_{\text{vac}}+\frac{1}{a_0}\Big(\theta\cdot\frac{\partial b_{\text{vac}}}{\partial\ln a}\Big)\frac{da_0}{d\theta}-\frac{N-1}{N}\,b_{\text{bulk}}, \end{equation}
  with \(b_{\text{vac}}\) and \(b_{\text{bulk}}\) the energy rows of the relaxed vacancy cell and the \(N\)-atom bulk cell (the bulk cell has zero virial at \(a_0\), so it
  has no \(a_0\) term).
\item
  \textbf{Elastic constants, relaxed-ion.} At each applied strain the internal coordinates \(u\) (atomic positions in the strained
  cell) are relaxed, and the stress \(\tau=-W/\Omega\) is linear in \(\theta\) at fixed geometry, \(\tau=B_\tau\theta\) with \(B_\tau=-B_V/\Omega\).
  Then
  \begin{gather}\frac{d\tau}{d\theta}=B_\tau+\frac{\partial\tau}{\partial u}\frac{du^*}{d\theta}+\frac{\partial\tau}{\partial a}\frac{da_0}{d\theta},\\
\frac{du^*}{d\theta}=-\mathcal H_{uu}^{-1}\Big[\Big(\frac{\partial b_E}{\partial u}\Big)^{\mathsf T}+\partial_a\nabla_uE\otimes\frac{da_0}{d\theta}\Big],
\end{gather}
  with \(\mathcal H_{uu}\) the Hessian in \(u\). The elastic constants are linear fits of \(\tau\) against strain, so their rows are the same fits of these. Stress route: the harness\textquotesingle s
  matscipy protocol (Voigt strain pattern \(e_1+e_4\), 5 steps of ±1 \%, linear fits). Energy route: curvature at the same steps
  of patterns \(e_1\), \(e_4\), \(e_1+e_2\).
\item
  \textbf{Decohesion.} Rigid openings of the relaxed bulk: linear in \(\theta\) at fixed \(a_0\) plus the \(a_0\) term; the cubic-spline
  derivative is linear in the energies.
\item
  \textbf{Discrete steps} (\(E(V)\)-grid argmin, maximum over a grid) are held at their index; \(\gamma_{\text{us}}\) and phonons are
  committee-only. So is the bcc phase energy: for both models its \(E(V)\) minimum is not at the grid point of the freely relaxed
  volume, so the selected point changes with \(\theta\) and has no derivative. The delta method therefore covers 17 of the 20 scored QoIs.
\end{itemize}

Rows are checked against finite differences in \(\theta\) (to \textasciitilde \(10^{-5}\)) and the values they imply reproduce the harness to
\textasciitilde \(10^{-10}\) (energy QoIs) and 0.04 GPa (elastic constants).

\subsubsection{2.7.3 Committees}\label{committees}

\(K=200\) draws \(\theta_k=\hat\theta+L\xi_k\), \(\xi_k\sim\mathcal N(0,I_r)\) (fixed seeds), are run through the unmodified benchmark harness. Per the protocol,
the draws are made once at \(\lambda=1\) and QoI deviations are rescaled, \(Q_k(\lambda)=\hat Q+\lambda(Q_k-\hat Q)\), with \(Q_k\) member \(k\)\textquotesingle s QoI and
\(\hat Q\) the mean model\textquotesingle s; a full
re-run at \(\lambda_F\) tests that linearisation. In that re-run a member is rejected, by a rule fixed beforehand that uses no
DFT value, if a test fails, if any relaxed interatomic distance falls below 2.0 Å (collapse), or if its diamond cell is
Born-unstable. The headline \(\sigma\) is the delta-method \(\sigma\) where it agrees with the \(\lambda=1\) committee to 10 \%, else the committee\textquotesingle s.

\protect\phantomsection\label{method-disc}
\subsubsection{2.7.4 Measuring the bias: a discrepancy term}\label{measuring-the-bias-a-discrepancy-term}

A bias shared by every training configuration of a kind is invisible to any delete-one jackknife: deleting one
configuration does not move the fit along it. The defect formation energies expose such a bias. Every training interstitial,
vacancy and divacancy cell is a hot molecular-dynamics snapshot, while the QoIs are relaxed defects.

\textbf{Measuring the bias from training labels.} For each defect class
\(t\in\{\text{interstitial},\text{vacancy},\text{divacancy}\}\), the bulk reference is a per-atom thermal trend
\(e_{\text{b}}(\eta)=\beta_0+\beta_1\eta\), fitted by least squares to the residuals per atom of the diamond bulk cells within
0.6 Å\textsuperscript{3}/atom of the defect cells\textquotesingle{} mean volume, with \(\eta=F_{\text{rms}}^2\) the mean squared DFT force component of
a cell (proportional to the thermal mean-square displacement for a harmonic solid). Each training defect cell \(i\), with
\(n_i\) atoms, energy row \(b_i\) and energy label \(y_i\), then gives a bulk-subtracted residual
\(u_i=y_i-b_i\cdot\hat\theta-n_i\,e_{\text{b}}(\eta_i)\). Regressing \(u_i\) on \(\eta_i\) and extrapolating to \(\eta\to0\)
estimates the model\textquotesingle s error for a relaxed defect. The intercept is linear in the labels and in \(\hat\theta\), so it is a
synthetic observation:
\begin{gather}r^*_t=\sum_i a_i u_i=y^*_t-d^*_t\cdot\hat\theta,\\
a=e_1^{\mathsf T}(X^{\mathsf T}X)^{-1}X^{\mathsf T},\\
X=[1,\,\eta_i],
\end{gather}
with \(X\) the regression design (rows \([1,\eta_i]\)), \(e_1=(1,0)^{\mathsf T}\) selecting the intercept, \(d^*_t\) and
\(y^*_t\) the same combination of the defect rows and labels with the bulk trend\textquotesingle s rows and labels subtracted, and standard
error \(s^*_t\) from the regression scatter. No QoI value or tolerance enters.

\textbf{Training-cell size.} The training vacancy cells come in two sizes, 63 and 215 atoms, so the vacancy trend need not be
shared. We compare four linear-Gaussian models of the vacancy residuals \(u_i\) by their evidence: one intercept and slope for
both sizes; per-size intercepts with a shared slope; a shared intercept with per-size slopes; and per-size intercepts and
slopes. The coefficients have Gaussian priors (intercepts \(\mathcal N(0,s_a^2)\) with \(s_a=1\) eV, slopes
scaled so that a slope moves \(u\) by about \(s_a\) across the data) and are integrated analytically. Each size has its
own noise level, log-uniform on {[}\(10^{-3}\), 10{]} eV and integrated on a grid. The QoI vacancy sits in a 216-site cell, so
its bias is the 215-atom intercept, averaged over the four models with their posterior probabilities (equal prior odds). This
uses training labels only, as for the interstitials.

\textbf{Discrepancy term.} For the QoIs of defect class \(t\), the variance becomes
\(\sigma_Q^2=\sigma_{Q,\text{method}}^2+r^{*2}_t+s^{*2}_t\), with \(\sigma_{Q,\text{method}}\) the \(\sigma\) of the uncertainty method without the term (interstitial class for the three interstitials; vacancy class for the
vacancy). Every other QoI is unchanged. This leaves the pure parameter-space propagation for these QoIs and is reported as a
separate, labelled method. The term widens \(\sigma\) but leaves the prediction at the potential\textquotesingle s own value, on purpose: the uncertainty
describes the potential that is run. Subtracting \(r^*_t\) from the prediction would correct the potential, a different object
and one that no longer has a parameter vector; Table~\ref{tab:14} gives the errors of that corrected prediction for reference.

Nothing in this analysis is specific to the CR3 propagation. The bias \(r^*_t\) and its standard error come from the
training labels and the fitted mean alone, so the same term can be added to the \(\sigma_Q\) of any method that shares that mean,
the baselines included.

\protect\phantomsection\label{method-valid}
\section{3 Results}\label{results}

\subsection{3.1 Validation statistics}\label{validation-statistics}

Section 2 ends with a \(\sigma\) for every prediction we test: a per-atom force covariance and conformal region (Section 2.5), a
scaled CR3 covariance in \(\theta\) (Section 2.4), and from it a \(\sigma_Q\) for each QoI by the delta method or by committees, with or
without the discrepancy term (Sections 2.7 and 2.7.4). The same few statistics judge all of them against held-out data or DFT. Two
held-out sets are used, and they play different roles. The per-atom quantile comes from a hold-out of the training set, so the
test split is an independent test of it. The propagation\textquotesingle s global scale and B1\textquotesingle s quantile are fitted on the test split\textquotesingle s
residuals, as the specification fixed (Section 3.3), so their per-atom coverage on the test split is in sample. No DFT QoI value
enters any scale.

\textbf{Sign conventions.} An error in Table~\ref{tab:1} and in \(z_Q\) below is \(\text{model}-\text{DFT}\). The defect analysis
(Section 3.8) works with training residuals, \(\text{DFT}-\text{model}\), and its tables say so.

For the per-atom forces (Section 3.2), the test is the coverage of the conformal regions on the 248 test configurations,
overall and per configuration type, together with the RMS of the observed force error over the predicted \(\sigma_{F,i}\), which is
1 for a calibrated Gaussian reading. For the QoIs (Sections 3.4--3.8), each prediction gives a standardised error
\(z_Q=(\hat Q-Q_{\text{DFT}})/\sigma_Q\), with \(\hat Q\) the mean model\textquotesingle s value. We count how many of the 20 scored QoIs fall inside
the Gaussian 68 \% and 95 \% intervals. Each count \(k\) of \(n\) gives a posterior for the coverage probability, \(\mathrm{Beta}(k+\tfrac12,\,n-k+\tfrac12)\) from
the Jeffreys prior, and we quote its 95 \% credible interval and the posterior probability that the coverage reaches its nominal value.
With only 20 QoIs these are broad: 17 of 20 at 95 \% gives a credible interval of {[}0.65, 0.96{]}, with \(P(\text{coverage}\ge0.95)=0.04.\)

Counts treat the QoIs as independent, but they share the same coefficients, so their errors are correlated. Section 3.5 therefore
scores the whole vector of DFT values at once, by the evidence each method\textquotesingle s predictive distribution gives it. With
\(\Delta\) the vector of \((Q_{\text{DFT}}-\hat Q)/\text{tol}_Q\) over the scored QoIs and \(C_Q\) its predicted covariance in the same
tolerance units (\(\lambda^2G\,LL^{\mathsf T}G^{\mathsf T}\) for the delta method, with \(G\) the QoI rows stacked, or the committee
covariance),
\begin{multline}\log p(\Delta\mid\text{method})=-\tfrac12\,\Delta^{\mathsf T}(C_Q+\epsilon I)^{-1}\Delta\\
-\tfrac12\log\lvert 2\pi(C_Q+\epsilon I)\rvert ,
\end{multline}
with a small ridge \(\epsilon=10^{-6}\) (a floor of \(10^{-3}\) tolerances on \(\sigma\); \(10^{-8}\) and \(10^{-4}\) as
sensitivities). Every method\textquotesingle s scale is fixed beforehand, on held-out forces or held-out scores, so no method has a free parameter left,
and the difference of two methods\textquotesingle{} log evidence on the same QoIs is a log Bayes factor. The first term rewards predictions whose
errors, and their correlations, match the covariance; the second penalises intervals that are wider than they need to be. The evidence
therefore weighs calibration and sharpness together, and a method cannot gain by inflating its intervals.

Coverage alone rewards intervals that are too wide. We therefore also report sharpness, the median of \(\sigma_Q/\text{tol}_Q\)
with \(\text{tol}_Q\) the benchmark\textquotesingle s tolerance for QoI \(Q\): a value below 1 means the interval is narrower than the accuracy the
benchmark asks for. A useful method has near-nominal coverage at the smallest sharpness. The CRPS combines the two in one number per QoI\cite{gneiting2007}. For a Gaussian prediction with mean
\(\hat Q\) and standard deviation \(\sigma_Q\) it is
\begin{equation} \mathrm{CRPS}=\sigma_Q\Big[z\big(2\Phi(z)-1\big)+2\varphi(z)-\frac{1}{\sqrt\pi}\Big], \end{equation}
with \(z\) the standardised error and \(\Phi\),
\(\varphi\) the standard normal distribution and density functions. It is a proper score: it is smallest in expectation for the
true predictive distribution, it grows for intervals that are too narrow and for intervals that are too wide, and it tends
to the absolute error as \(\sigma_Q\) → 0. We report its mean over the scored QoIs in tolerance units, CRPS/tol (lower is
better); committee methods are scored with the Gaussian of their spread, as for coverage.

\protect\phantomsection\label{results-forces}
\subsection{3.2 Calibrated per-atom forces}\label{calibrated-per-atom-forces}

\begin{figure*}[tbp]
\centering
\includegraphics[width=\textwidth,height=0.6\textheight,keepaspectratio]{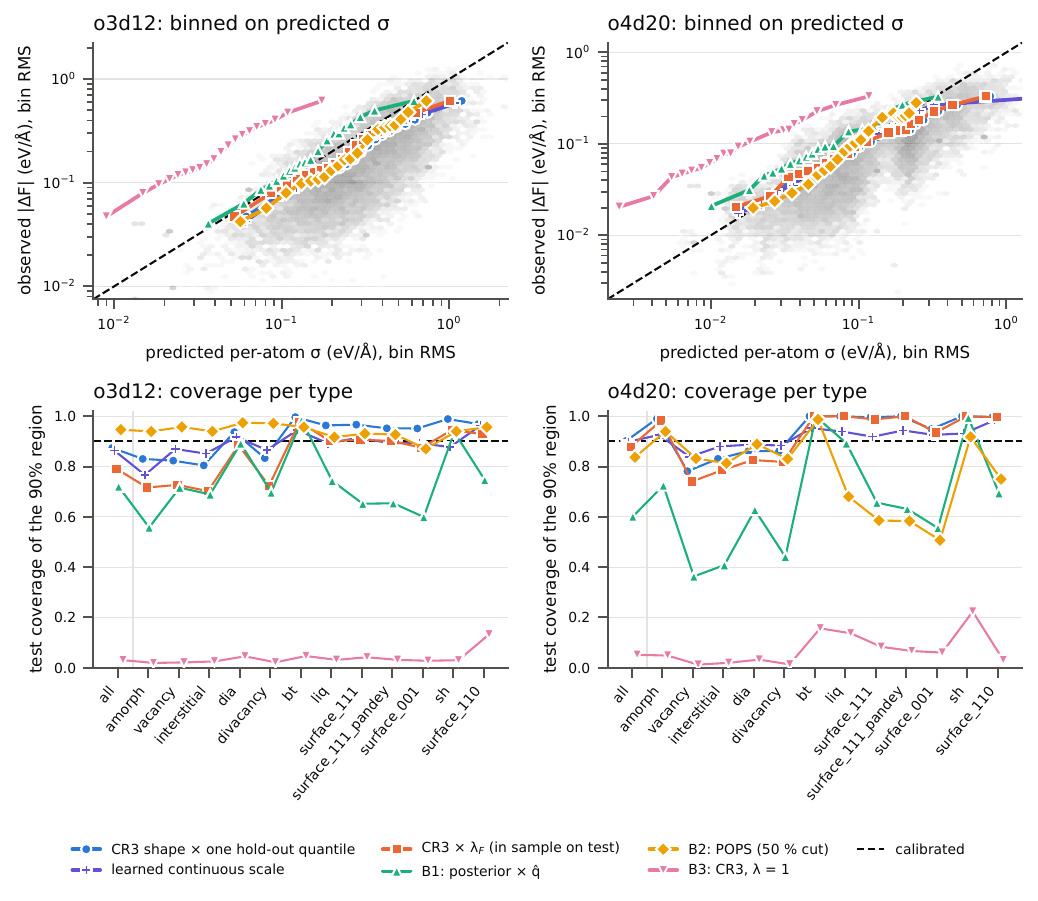}
\caption{Top: binned on predicted \(\sigma\) (20 equal-count bins), RMS observed \(|\Delta F|\) against RMS \(\sigma\), for the CR3 shape with one hold-out
quantile (the method) and with the learned scale (\(\sigma_{F,i}\) from the 90 \% quantile), and for the per-atom \(\sigma\) of the
propagation\textquotesingle s methods: CR3 × \(\lambda_F\) (the headline), B1 (posterior × \(\hat q\)), B2 (POPS, 50 \% leverage cut; the
box\textquotesingle s uniform variance) and B3 (CR3, \(\lambda=1\)); the per-atom density behind. Points below the diagonal are conservative.
Bottom: test coverage of the 90 \% region, overall and per GAP-18 configuration type with at least 300 test atoms. Every
method\textquotesingle s region is the same kind: the ellipsoid of its own 3×3 force covariance, \(e^{\mathsf T}C^{-1}e\le\chi^2_3(0.9)\)
(for the one quantile and the learned scale this is their conformal region). The headline\textquotesingle s \(\lambda_F\) is fitted on the test split itself, so its per-atom coverage here is in sample.}\label{fig:3}
\end{figure*}

Figure~\ref{fig:3} and Table~\ref{tab:3} test the per-atom force uncertainty of Section 2.5 on the 248 test configurations, with the BLR fits
whose CR3 posteriors the QoIs use; atoms whose force rows vanish by site symmetry are dropped by the rule of Section 3.3. At a
nominal 90 \%, the CR3 shape with one hold-out quantile covers 0.87 of the test atoms for o3d12 and
0.90 for o4d20. o3d12 is below nominal, and the cause is the transfer from the hold-out fit to the full fit. The calibration scores come from the hold-out posterior,
fitted on 1782 of the 2227 training configurations, and are carried to the full posterior by
\((N/N_{\text{fit}})^{\beta}\), with \(\beta\) estimated from a second, smaller hold-out fit (1427 configurations) and
clipped to {[}0, 1/2{]} (Section 2.5). For o4d20 that estimate gave \(\beta=0.43\), a factor 1.101, and the
test scores\textquotesingle{} 90 \% quantile is 0.991 times the transferred one. For o3d12 the scale of the
smaller fit came out larger (8.56 against 7.03), so \(\beta_{\text{raw}}=-0.88\) was
clipped to 0 and nothing was transferred; the full fit\textquotesingle s test scores are 6.8 \% larger at the 90 \% quantile, which is
the shortfall. We report this as found: the transfer rule was fixed before any test coverage was seen, and changing it now would
be tuning on the test split. Where the transfer worked, for o4d20, one quantile reaches the nominal level overall with no
environment-dependent scale: the shape carries most of the dependence of the error on the environment. The estimate of \(\beta\) is
fragile in one respect. It is taken from RMS scales, which a few atoms with almost no CR3 variance can dominate: in the
out-of-distribution refits below, the first group\textquotesingle s RMS scale is \(3.6\times10^{9}\) (o4d20) where its
90 \% quantile is 17.2, and that refit\textquotesingle s \(\beta=0.23\) (factor
1.053) reflects those atoms rather than the data size. An estimate from quantiles would be robust to
them; as with the clipping, we keep the rule as it was fixed.

The coverage is not even across environments, though. The point-defect cells are under-covered and the liquid and surface
cells over-covered (Figure~\ref{fig:3}, bottom). The amorphous cells differ between the models: under-covered for o3d12
(0.83) and over-covered for o4d20 (0.99).
The learned scale of Section 2.5 corrects much of this in distribution: it covers 0.87 and 0.91
overall (ranges over the five hold-out splits 0.86--0.88 and 0.90--0.91), with similar median region sizes
(17.2 against 17.1 for o3d12, 15.0 against
16.6 for o4d20, in units of the score), and the RMS deviation of the per-type coverage from 0.90 falls
from 0.072 to 0.044 for o3d12 and from 0.087 to
0.041 for o4d20 (Table~\ref{tab:3}; for o3d12 the amorphous cells are the exception). Out of distribution it is
less robust, as the next test shows.

How the quantile weights the calibration atoms matters more than either scale. A quantile that counts each calibration
configuration once in total is about 12 \% smaller and covers only 0.82 and
0.84, because the many small, well-predicted cells then dominate it; the quantiles here are
atom-weighted. The same CR3
shape with the propagation\textquotesingle s global \(\lambda_F\) covers 0.79 and 0.88, below the
one quantile for both models and with the same pattern across configuration types, even though \(\lambda_F\) is fitted on these test
configurations (in sample). \(\lambda_F\) is an RMS over force components, not a quantile of the 3-D score, and the CR3
shape is strongly anisotropic, so errors along its thin directions leave the ellipsoid. All regions in Figure~\ref{fig:3} are
ellipsoids of the method\textquotesingle s own 3×3 covariance; an isotropic ball \(|\Delta F|\le\sigma\,(\chi^2_3(0.9)/3)^{1/2}\) of the same
\(\sigma=(\operatorname{tr}C)^{1/2}\) is larger along those directions and covers more (0.95 and 0.89 for
\(\lambda_F\)); compared that way, \(\lambda_F\) would look better calibrated than the one quantile, but only because its region is larger. Increasing the regulariser \(\varepsilon\) of the ellipsoid tenfold changes no coverage by more than 0.012.

\begin{table*}[htbp]
\centering\small\setlength{\tabcolsep}{3pt}
\caption{Test coverage of the 90 \% per-atom force region per GAP-18 configuration type (types
with at least 300 test atoms): one hold-out quantile against the learned scale (mean over five hold-out splits), and the RMS
deviation of the per-type coverage from nominal; and the test-split scales of the propagation, \(\lambda_F\) and
\(\lambda_V\) = RMS \(z\) with the CR3 shape over that type\textquotesingle s force and virial rows (Section 3.3; "--": no virials).}\label{tab:3}
\fitwidth{\begin{tabular}{@{}llllllll@{}}
\toprule
model & configuration type & atoms & configs & one quantile & learned scale & \(\lambda_F\) & \(\lambda_V\) \\
\midrule
o3d12 & amorph & 3304 & 16 & 0.83 & 0.80 & 6.4 & -- \\
o3d12 & vacancy & 2691 & 21 & 0.82 & 0.88 & 6.0 & 4.9 \\
o3d12 & interstitial & 2543 & 12 & 0.80 & 0.86 & 7.5 & 3.8 \\
o3d12 & dia & 2144 & 49 & 0.94 & 0.91 & 4.1 & 4.2 \\
o3d12 & divacancy & 1712 & 8 & 0.83 & 0.88 & 5.8 & -- \\
o3d12 & bt & 1546 & 44 & 1.00 & 0.94 & 3.8 & 14.6 \\
o3d12 & liq & 576 & 8 & 0.96 & 0.92 & 5.0 & 4.4 \\
o3d12 & surface\_111 & 480 & 5 & 0.97 & 0.93 & 5.2 & -- \\
o3d12 & surface\_111\_pandey & 480 & 5 & 0.95 & 0.91 & 5.4 & -- \\
o3d12 & surface\_001 & 432 & 3 & 0.95 & 0.90 & 5.5 & -- \\
o3d12 & sh & 367 & 13 & 0.99 & 0.92 & 4.1 & 4.4 \\
o3d12 & surface\_110 & 324 & 3 & 0.97 & 0.97 & 4.8 & -- \\
o3d12 & RMS deviation from 0.90 & & & 0.072 & 0.044 & & \\
o4d20 & amorph & 3304 & 16 & 0.99 & 0.93 & 4.5 & -- \\
o4d20 & vacancy & 2691 & 21 & 0.78 & 0.85 & 8.2 & 5.2 \\
o4d20 & interstitial & 2543 & 12 & 0.83 & 0.89 & 8.0 & 5.4 \\
o4d20 & dia & 2144 & 49 & 0.86 & 0.90 & 6.5 & 3.7 \\
o4d20 & divacancy & 1712 & 8 & 0.86 & 0.89 & 6.9 & -- \\
o4d20 & bt & 1546 & 44 & 1.00 & 0.95 & 2.7 & 15.5 \\
o4d20 & liq & 576 & 8 & 1.00 & 0.94 & 3.0 & 2.5 \\
o4d20 & surface\_111 & 480 & 5 & 1.00 & 0.92 & 4.0 & -- \\
o4d20 & surface\_111\_pandey & 480 & 5 & 1.00 & 0.93 & 3.9 & -- \\
o4d20 & surface\_001 & 432 & 3 & 0.95 & 0.94 & 5.0 & -- \\
o4d20 & sh & 367 & 13 & 1.00 & 0.94 & 2.5 & 2.2 \\
o4d20 & surface\_110 & 324 & 3 & 1.00 & 0.99 & 3.9 & -- \\
o4d20 & RMS deviation from 0.90 & & & 0.087 & 0.041 & & \\
\bottomrule
\end{tabular}}
\end{table*}

An out-of-distribution test asks whether this holds for environments absent from the fit. We refitted both models (BLR,
evidence-optimised on the reduced training set) with GAP-18\textquotesingle s cracks ((111) and (110) cleavage, 1-10 front) and screw dislocations removed from training, from the
calibration hold-out and from the test file, and scored the 6056 atoms in 36 configurations that were removed (Table~\ref{tab:4}). The design follows
the big-cell check of an alloy model in the \texttt{ace-jax} development tests\cite{acejax_cantor}. The global scale is fitted on the
in-distribution test split, so it is out of sample here, like the conformal ones. Over all held-out atoms one quantile covers
0.86 (o3d12) and 0.93 (o4d20), and the global scale 0.63 and
0.89. The cracks are covered ((111) 0.91 and (110)
0.95 for o3d12; 0.91 and
0.96 for o4d20). The screw dislocations are under-covered for o3d12
(0.77) but not for o4d20 (0.91).

Out of distribution the learned scale gains nothing, and it can lose. For o3d12 it covers 0.86 of the held-out atoms,
as one quantile does (0.86), and the two agree to 0.01 on each held-out type (Table~\ref{tab:4}); for o4d20 it
covers 0.87 against one quantile\textquotesingle s 0.93, and it falls below nominal on the (111)
crack (0.88 against 0.91) and on the screw
dislocations (0.84 against 0.91), which one quantile
covers. The earlier ARD fit of o4d20 showed the same pattern. The risk is structural: a learned scale narrows the regions of environments whose training
counterparts the model reproduces well, mostly near-bulk ones, and atoms at dislocation cores and on crack flanks resemble those
in descriptor space but can carry larger errors. One quantile cannot narrow selectively, and it is simpler, which is why it is the
method here; the learned scale is a refinement for target environments that are represented in the training data.

Ho, Ortner and Wang\cite{ho2026} reach a more favourable verdict on their learned scale out of distribution, on compositions and
exchange-correlation functionals outside the calibration set. Their test is the correlation of the predicted uncertainty with the
error; ours is coverage. These are different properties\cite{shanks2026}. The two results are not in conflict, since a scale can rank errors better and still be too narrow where the model
extrapolates, but they show that a learned scale should be checked for coverage on environments left out of its calibration.

\begin{table*}[htbp]
\centering\small\setlength{\tabcolsep}{3pt}
\caption{Out-of-distribution per-atom force coverage at nominal 90 \%: BLR refits without
cracks and screw dislocations, scored on them (90 \% intervals from a bootstrap over configurations). "ID test": the same
refit\textquotesingle s coverage on the in-distribution test split.}\label{tab:4}
\fitwidth{\begin{tabular}{@{}lllllll@{}}
\toprule
model & method & ID test & all held out & crack (110) & crack (111) & screw dislocation \\
\midrule
o3d12 & one hold-out quantile & 0.88 & 0.86 {[}0.83, 0.88{]} & 0.95 {[}0.94, 0.96{]} & 0.91 {[}0.90, 0.92{]} & 0.77 {[}0.76, 0.78{]} \\
o3d12 & learned scale & 0.88 & 0.86 {[}0.83, 0.87{]} & 0.94 {[}0.93, 0.95{]} & 0.90 {[}0.88, 0.90{]} & 0.78 {[}0.76, 0.79{]} \\
o3d12 & global scale (ID test) & 0.67 & 0.63 {[}0.59, 0.66{]} & 0.78 {[}0.76, 0.80{]} & 0.69 {[}0.67, 0.72{]} & 0.50 {[}0.49, 0.52{]} \\
o4d20 & one hold-out quantile & 0.90 & 0.93 {[}0.91, 0.94{]} & 0.96 {[}0.96, 0.97{]} & 0.91 {[}0.90, 0.93{]} & 0.91 {[}0.90, 0.93{]} \\
o4d20 & learned scale & 0.89 & 0.87 {[}0.84, 0.88{]} & 0.93 {[}0.91, 0.94{]} & 0.88 {[}0.84, 0.89{]} & 0.84 {[}0.80, 0.84{]} \\
o4d20 & global scale (ID test) & 0.86 & 0.89 {[}0.88, 0.91{]} & 0.94 {[}0.93, 0.96{]} & 0.88 {[}0.87, 0.90{]} & 0.87 {[}0.85, 0.89{]} \\
\bottomrule
\end{tabular}}
\end{table*}

Figure~\ref{fig:3} also shows the per-atom forces of the baselines, to which the QoI comparison of Section 3.4 applies. B3 is
about \(\lambda_F\) times too narrow everywhere. B1, the posterior shape with one scale, under-covers
(0.72 for o3d12, 0.60 for o4d20), most in the amorphous, point-defect and surface cells. B2
over-covers for o3d12 (0.95; per type 0.87 (surface\_001) to 0.97). For o4d20 it under-covers overall
(0.84) and unevenly: per type 0.51 (surface\_001) to 0.99, low on the liquid and surface cells.

\protect\phantomsection\label{method-calib}
\subsection{3.3 One global scale for propagation}\label{one-global-scale-for-propagation}

The per-atom scale of Section 3.2 calibrates force outputs atom by atom, but it does not define a covariance in \(\theta\)
(Section 2.5). Propagation needs a single scale \(\lambda\) for the CR3 distribution \(\theta\sim\mathcal N(\hat\theta,\lambda^2LL^{\mathsf T})\).
We fit it on the same 248 held-out test configurations as Section 3.2, with the same fit and CR3 shape, but over all rows of a
type at once. Every row \(b\) gives a standardised score
\(z=(y-\hat y)/\lVert L^{\mathsf T}b\rVert\), with \(\hat y=b\cdot\hat\theta\) (energies per atom, force components, virial components). The
scale for observable type \(q\in\{E,F,V\}\) is \(\lambda_q=\operatorname{RMS}(z_q)\) over that type\textquotesingle s rows; \(\lambda_{\text{pool}}\) weights E, F
and V equally within a configuration and configurations equally. As in Section 3.2, rows that vanish by site symmetry
(\(\lVert L^{\mathsf T}b\rVert\lt 10^{-3}\) of the median of their kind) are dropped. No QoI, and no DFT QoI value, enters any scale.
The headline uses \(\lambda_F\): forces are most of the training rows, and they are the quantity calibrated per atom in Section 3.2.
In Figure~\ref{fig:3}, \(\lambda_F\) gives the headline line.

A single scale is a compromise. Table~\ref{tab:3} gives the test-split scales per configuration type. Over the types listed
there \(\lambda_F\) ranges from 3.8 to 7.5 (o3d12) and 2.5 to 8.2 (o4d20), and
\(\lambda_V\) is largest for the bt cells (14.6 and 15.5). The ratios of the global scales (Table~\ref{tab:5}) show the same at the level of observables.

\begin{table*}[htbp]
\centering\small\setlength{\tabcolsep}{3pt}
\caption{Test-split scales (CR3 shape).}\label{tab:5}
\fitwidth{\begin{tabular}{@{}lllllll@{}}
\toprule
model & \(\lambda_E\) & \(\lambda_F\) & \(\lambda_V\) & \(\lambda_{\text{pool}}\) & \(\lambda_V/\lambda_F\) & \(\lambda_E/\lambda_F\) \\
\midrule
o3d12 & 4.92 & 5.82 & 8.03 & 5.85 & 1.38 & 0.84 \\
o4d20 & 5.03 & 6.17 & 8.23 & 6.05 & 1.33 & 0.82 \\
\bottomrule
\end{tabular}}
\end{table*}

Whether the observables need different scales is a model comparison. The test configurations are the independent
units, as in CR3, because the force components of one configuration are correlated. For configuration \(c\) and observable \(o\), let
\(m_{co}\) be the mean of \(z^2\) over the block\textquotesingle s components (1 energy, \(3N\) forces, 6 virial components). We model
\(m_{co}=\lambda_o^2X\) with \(X\sim\chi^2_{\nu_o}/\nu_o\), a Gamma distribution with mean \(\lambda_o^2\), as the RMS-z
definition of \(\lambda\) requires. The effective number of degrees of freedom \(\nu_o\) absorbs the correlation within a block and the
heavy tails. It is free for each observable in every model, log-uniform on {[}0.05, 1000{]}, and integrated out. The models are
the five ways to let E, F and V share scales, each scale log-uniform on {[}0.1, 1000{]}. A joint Gaussian likelihood of each block
under \(\lambda^2\) times the full CR3 block covariance was also tried and is not usable: for forces it is set by the regularisation of a
nearly singular block, that is by the off-diagonal CR3 shape, not by the scale.

\begin{table}[htbp]
\centering\small\setlength{\tabcolsep}{3pt}
\caption{Test-split evidence for shared against per-observable scales: log Bayes factor of
each way of sharing scales against the best (configurations as units, equal prior odds).}\label{tab:6}
\fitwidth{\begin{tabular}{@{}lll@{}}
\toprule
scales & o3d12 & o4d20 \\
\midrule
one scale & -31.6 & -23.8 \\
E \textbar{} F = V & -32.7 & -23.7 \\
F \textbar{} E = V & -22.3 & -19.4 \\
V \textbar{} E = F & +0.0 & +0.0 \\
E \textbar{} F \textbar{} V & -4.2 & -3.9 \\
\bottomrule
\end{tabular}}
\end{table}

In both models the evidence favours one scale for energies and forces and a separate, larger one for virials:
log Bayes factors of 32 (o3d12) and 24 (o4d20) against one scale for all, and
4.2 and 3.9 against three separate scales. The posterior scales are \(\lambda_V\) 8.1 {[}7.2, 9.0{]}, \(\lambda_{E=F}\) 5.0 {[}4.8, 5.2{]} for
o3d12 and \(\lambda_V\) 8.3 {[}7.3, 9.3{]}, \(\lambda_{E=F}\) 5.2 {[}4.9, 5.5{]} for o4d20 (mean and 95 \% credible interval). With each configuration weighted equally,
\(\lambda_E\) and \(\lambda_F\) agree; the smaller \(\lambda_E\) of Table~\ref{tab:5} comes from the component-weighted RMS, in which
the largest cells dominate \(\lambda_F\). The degree of misspecification therefore depends on the
observable, and with \(\lambda_F\) virials are under-covered on the test split. The QoIs most like virials are the elastic
constants, which are derivatives of the stress, and the phase energies, which come from energy--volume curves, so their
intervals at \(\lambda_F\) may be too narrow; Section 3.4 reports the coverage with \(\lambda_V\) as a sensitivity.

\protect\phantomsection\label{results}
\subsection{3.4 Coverage and sharpness}\label{coverage-and-sharpness}

\begin{table*}[htbp]
\centering\small\setlength{\tabcolsep}{3pt}
\caption{Coverage of the DFT values (count {[}95 \% credible interval of the coverage probability{]}) and sharpness, for the four methods of Figures~\ref{fig:5} and \ref{fig:6}.}\label{tab:7}
\fitwidth{\begin{tabular}{@{}llllllll@{}}
\toprule
model & method & \(n\) & 68 \% & 95 \% & RMS \(z\) & median \(\sigma/\text{tol}\) & CRPS/tol \\
\midrule
o3d12 & CR3 × \(\lambda_F\) (headline) & 20 & 14 {[}0.48, 0.86{]} & 17 {[}0.65, 0.96{]} & 1.29 & 2.52 & 2.79 \\
o3d12 & B1: posterior × \(\hat q\) & 20 & 13 {[}0.43, 0.83{]} & 15 {[}0.54, 0.90{]} & 2.28 & 1.63 & 2.92 \\
o3d12 & B2: POPS (leverage cut 50 \%, the package default) & 20 & 13 {[}0.43, 0.83{]} & 15 {[}0.54, 0.90{]} & 1.55 & 1.75 & 2.86 \\
o3d12 & B3: CR3, \(\lambda=1\) & 20 & 5 {[}0.10, 0.46{]} & 8 {[}0.21, 0.62{]} & 7.53 & 0.43 & 3.16 \\
o4d20 & CR3 × \(\lambda_F\) (headline) & 20 & 12 {[}0.38, 0.79{]} & 15 {[}0.54, 0.90{]} & 1.31 & 0.64 & 0.75 \\
o4d20 & B1: posterior × \(\hat q\) & 20 & 11 {[}0.34, 0.75{]} & 12 {[}0.38, 0.79{]} & 2.58 & 0.37 & 0.82 \\
o4d20 & B2: POPS (leverage cut 50 \%, the package default) & 20 & 11 {[}0.34, 0.75{]} & 17 {[}0.65, 0.96{]} & 1.69 & 0.60 & 0.81 \\
o4d20 & B3: CR3, \(\lambda=1\) & 20 & 2 {[}0.02, 0.28{]} & 5 {[}0.10, 0.46{]} & 8.11 & 0.10 & 0.91 \\
\bottomrule
\end{tabular}}
\end{table*}

\begin{table*}[htbp]
\centering\small\setlength{\tabcolsep}{3pt}
\caption{Sensitivity of the coverage to the scale and to the committee variant: the CR3 shape
with the energy, virial or pooled scale (Section 3.3), the full re-run at \(\lambda_F\) keeping physical members only (Section 3.6),
and POPS with the 5 \% leverage cut instead of B2\textquotesingle s 50 \% (Section 3.7).}\label{tab:8}
\fitwidth{\begin{tabular}{@{}llllllll@{}}
\toprule
model & method & \(n\) & 68 \% & 95 \% & RMS \(z\) & median \(\sigma/\text{tol}\) & CRPS/tol \\
\midrule
o3d12 & CR3 × \(\lambda_V\) & 20 & 15 {[}0.54, 0.90{]} & 18 {[}0.72, 0.98{]} & 0.94 & 3.47 & 2.78 \\
o3d12 & CR3 × \(\lambda_E\) & 20 & 14 {[}0.48, 0.86{]} & 16 {[}0.59, 0.93{]} & 1.53 & 2.13 & 2.82 \\
o3d12 & CR3 × \(\lambda_{\text{pool}}\) & 20 & 14 {[}0.48, 0.86{]} & 17 {[}0.65, 0.96{]} & 1.29 & 2.53 & 2.79 \\
o3d12 & CR3 at \(\lambda_F\), full re-run, physical members & 20 & 14 {[}0.48, 0.86{]} & 15 {[}0.54, 0.90{]} & 1.30 & 4.49 & 5.12 \\
o3d12 & POPS, leverage cut 5 \% & 20 & 13 {[}0.43, 0.83{]} & 15 {[}0.54, 0.90{]} & 1.42 & 2.44 & 2.87 \\
o4d20 & CR3 × \(\lambda_V\) & 20 & 14 {[}0.48, 0.86{]} & 19 {[}0.79, 0.99{]} & 0.99 & 0.85 & 0.73 \\
o4d20 & CR3 × \(\lambda_E\) & 20 & 11 {[}0.34, 0.75{]} & 15 {[}0.54, 0.90{]} & 1.61 & 0.52 & 0.77 \\
o4d20 & CR3 × \(\lambda_{\text{pool}}\) & 20 & 12 {[}0.38, 0.79{]} & 15 {[}0.54, 0.90{]} & 1.34 & 0.63 & 0.75 \\
o4d20 & CR3 at \(\lambda_F\), full re-run, physical members & 20 & 15 {[}0.54, 0.90{]} & 17 {[}0.65, 0.96{]} & 1.06 & 1.01 & 1.41 \\
o4d20 & POPS, leverage cut 5 \% & 20 & 14 {[}0.48, 0.86{]} & 18 {[}0.72, 0.98{]} & 1.11 & 0.99 & 0.84 \\
\bottomrule
\end{tabular}}
\end{table*}

\begin{figure*}[tbp]
\centering
\includegraphics[width=\textwidth,height=0.8\textheight,keepaspectratio]{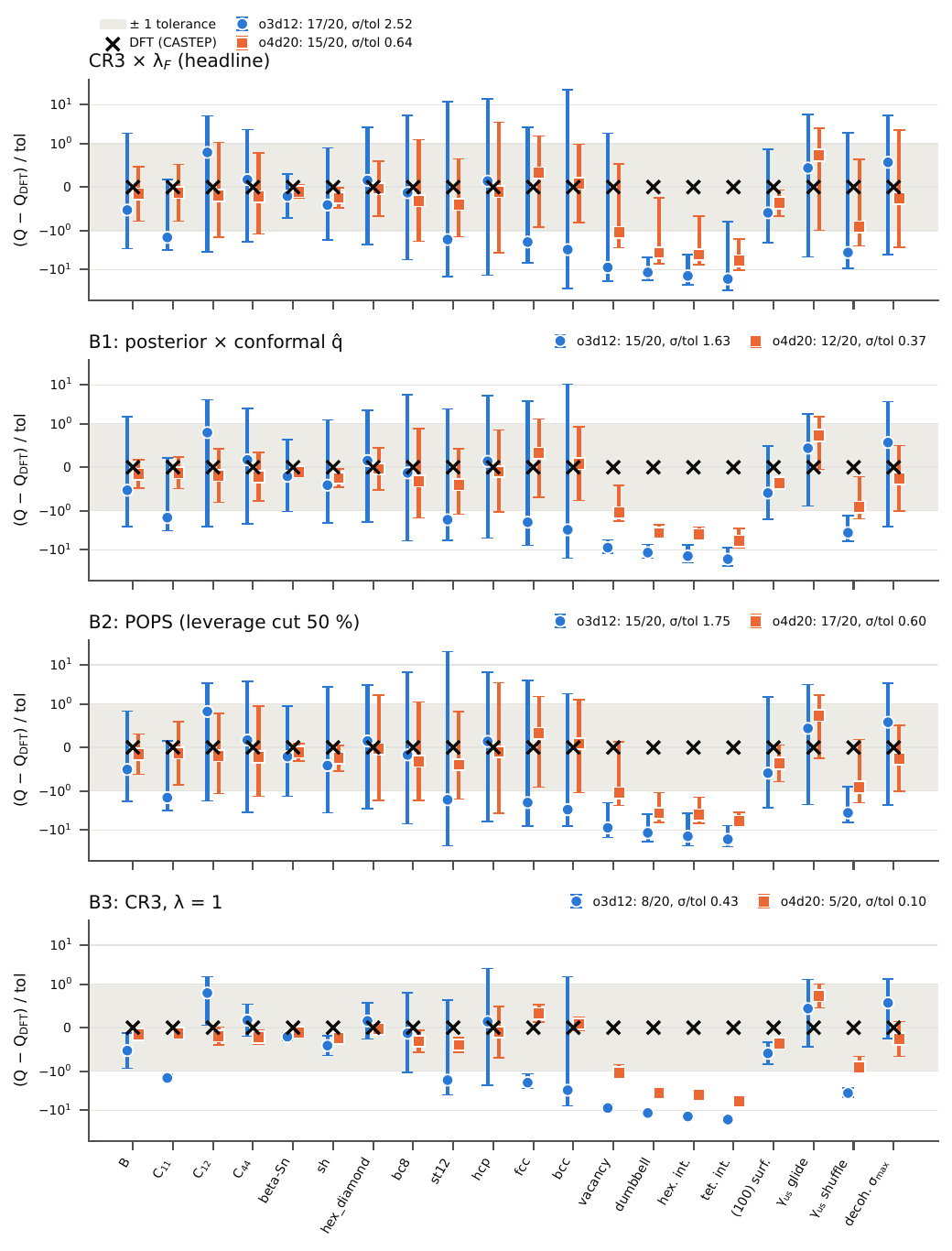}
\caption{Model prediction minus DFT, in units of the benchmark tolerance, with the 95 \% interval, for o3d12
(circles) and o4d20 (squares); DFT is the cross at 0 and the shaded band is ± 1 tolerance. The axis is linear within ± 1 and
logarithmic beyond, and is shared by all rows. Top row: the headline, CR3 × \(\lambda_F\); the larger model\textquotesingle s intervals are
3.9× tighter in median \(\sigma/\text{tol}\) while its coverage stays comparable. Below it the baselines of Section 2.6: B1, the posterior
scaled by one conformal quantile; B2, POPS with the 50 \% leverage cut (the reference package\textquotesingle s default); B3, CR3 without calibration (\(\lambda=1\)). Each row\textquotesingle s legend gives
its 95 \% coverage and median \(\sigma/\text{tol}\).}\label{fig:5}
\end{figure*}

\begin{figure*}[tbp]
\centering
\includegraphics[width=\textwidth,height=0.6\textheight,keepaspectratio]{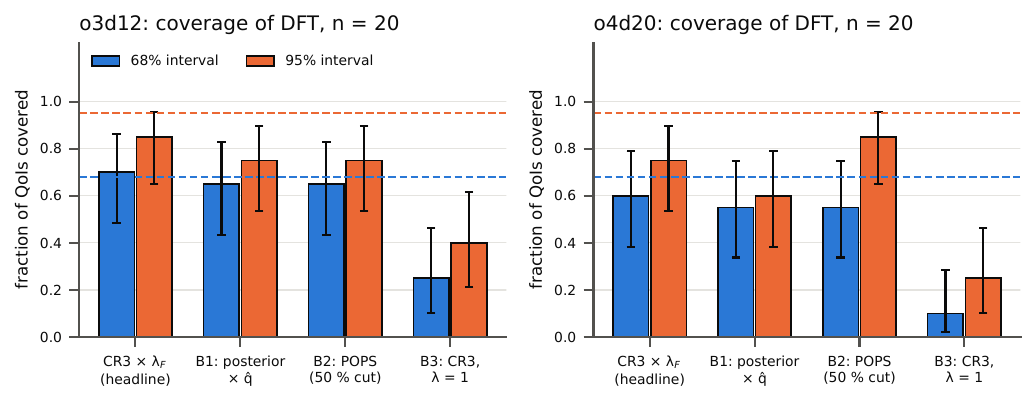}
\caption{Fraction of the 20 scored QoIs inside the 68 \% and 95 \% intervals for the four methods of Figure~\ref{fig:5}
(headline, B1, B2 with the 50 \% leverage cut, B3), with 95 \% credible intervals from the Jeffreys-prior Beta posterior; dashed lines are the nominal levels.}\label{fig:6}
\end{figure*}

With the headline CR3 × \(\lambda_F\), 17/20 QoIs of o3d12 and 15/20
of o4d20 lie inside their 95 \% intervals (Table~\ref{tab:7}, Figure~\ref{fig:5} top row, Figure~\ref{fig:6}). Both counts are below the nominal 19, and the coverage posteriors (Section 3.1) give
\(P(\text{coverage}\ge0.95)=0.04\) for o3d12 and 0.00 for o4d20: the headline under-covers at the 95 \% level. The misses are not spread across the QoIs. For o3d12 they are
the three interstitial formation energies; for o4d20 they are the three interstitial formation energies, the sh phase energy and the (100) surface energy. The interstitial formation energies, missed by both
models, are the subject of Section 3.8: they share a bias that a delete-one jackknife cannot see, not a scale that is too small.
Elsewhere the 95 \% intervals reach DFT, and the 68 \% counts (14 and
12 of 20, against a nominal 13.6) are close to nominal.

The intervals follow the accuracy of the model. The median \(\sigma/\text{tol}\) is 2.52 for o3d12 and
0.64 for o4d20, 3.9× narrower: o3d12\textquotesingle s intervals are wider than the benchmark tolerance, as they
should be for a model whose typical QoI error is about one tolerance (Figure~\ref{fig:1}), and o4d20\textquotesingle s are mostly inside it. So one scale fitted
to held-out forces gives similar coverage for both models while the intervals shrink with model quality. The CRPS, which weighs
calibration and sharpness in one score, ranks the headline first among the four methods for both models:
2.79 for o3d12 and 0.75 for o4d20, against
2.92 / 0.82 for B1,
2.86 / 0.81 for B2 and
3.16 / 0.91 for B3 (o3d12 / o4d20; Table~\ref{tab:7}). The margins are small
because the mean is dominated by the QoIs with the largest errors, which no width can fix: for o3d12 the three interstitials
are 12--17 tolerances off, so its CRPS/tol is near 3 for every method.

Of the baselines (Section 3.7), B1 and B3 cover fewer QoIs. POPS (B2) covers 15/20 (o3d12)
and 17/20 (o4d20), at widths close to the headline\textquotesingle s (median \(\sigma/\text{tol}\)
1.75 and 0.60). The sensitivity table (Table~\ref{tab:8}) shows that the choice of
scale matters at the margin. The virial scale \(\lambda_V\), larger than \(\lambda_F\) by about a third, covers
18/20 (o3d12) and 19/20 (o4d20) at the price of wider intervals
(median \(\sigma/\text{tol}\) 3.47 for o3d12 and 0.85 for o4d20); \(\lambda_E\) and \(\lambda_{\text{pool}}\) give about the
same as \(\lambda_F\). The full committee re-run at \(\lambda_F\), keeping only physical members, covers
15/20 for o3d12 and 17/20 for o4d20,
against 17/20 and 15/20 for the headline. The differences come mainly from the
phase energies, which are strongly nonlinear in \(\theta\) at \(\lambda_F\) (Section 3.6).

\begin{figure*}[tbp]
\centering
\includegraphics[width=\textwidth,height=0.6\textheight,keepaspectratio]{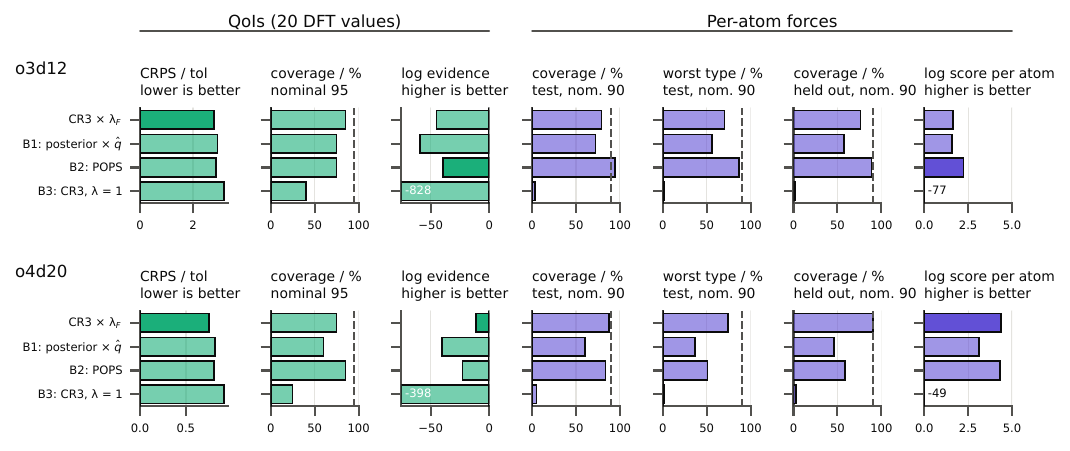}
\caption{The headline method, CR3 × \(\lambda_F\), and the three baselines: B1, the Bayesian posterior scaled by one conformal
quantile \(\hat q\); B2, POPS with the 50 \% leverage cut; B3, CR3 without calibration (\(\lambda=1\)). One row of panels per model. Left: the 20 scored QoIs, CRPS/tol, coverage of the 95 \% intervals, and the log evidence of the DFT values
(committee covariance, \(\varepsilon=10^{-6}\); Table~\ref{tab:9}). Right: per-atom forces, coverage of each method\textquotesingle s 90 \% ellipsoid on the test split,
its lowest value over the configuration types of Figure~\ref{fig:3} (Table~\ref{tab:3}), and on the cracks and screw dislocations held out
of BLR refits (Table~\ref{tab:4}), and the mean Gaussian log score per test atom.
Dashed: nominal coverage. Full colour: the best method on each proper score (CRPS, log evidence, log score); coverage panels are not ranked. On
the test split the headline\textquotesingle s \(\lambda_F\) and B1\textquotesingle s \(\hat q\) are fitted on the same data, so their test-split force panels are in sample. The log axes are cut at −75 (QoIs) and 0 (forces) so that the other methods can be
compared; B3\textquotesingle s values beyond them are written on the panel.}\label{fig:4}
\end{figure*}

Figure~\ref{fig:4} compares the four methods on both models. On the QoIs the headline has the lowest CRPS/tol for both models
and the highest log evidence for o4d20; for o3d12, B2 has the higher evidence (Table~\ref{tab:9}). The per-atom forces rank the
methods differently. B2\textquotesingle s box, read as a Gaussian of its covariance, covers 0.95 and 0.84 of the test atoms
at 90 \% (o3d12, o4d20), against 0.79 and 0.88 for the headline. On the held-out cracks and
dislocations the two models disagree. For o3d12 B2 covers 0.89 against the headline\textquotesingle s 0.76, and is higher on every
held-out type. For o4d20 it covers 0.59 against 0.91: 0.50 and 0.40 on the two cracks
(headline 0.95 and 0.89) and 0.77 on the screw dislocations (0.89), with the POPS ridge
selected as in distribution. The held-out mean log scores follow the coverages: 1.2 against 0.7 nats for o3d12, 1.0 against 2.5 for o4d20.
On the test split B2\textquotesingle s mean log score per atom
is 2.2 and 4.3 nats, against the headline\textquotesingle s 1.6 and 4.4. B1 under-covers the forces
in and out of distribution (0.72/0.58 and 0.60/0.46), and B3 covers almost no atoms. In distribution B2 is also uneven per configuration type for o4d20: its lowest per-type test coverage is 0.51, against 0.74 for the headline (o3d12: 0.87 and 0.70). The headline\textquotesingle s
\(\lambda_F\) and B1\textquotesingle s \(\hat q\) are fitted on the test split, so their test-split force scores are in sample; on the held-out
configurations every method is out of sample, with \(\lambda_F\), \(\hat q\) and the POPS box refitted on the reduced training set by
the same rules. The headline\textquotesingle s \(\lambda_F\) is the RMS over force components of Section 3.3; the global scale of Table~\ref{tab:4} is
fitted per atom instead (\(\lambda^2\) the mean of \(|\Delta F|\)\textsuperscript{2}/tr V), so its coverages differ.

\subsection{3.5 Joint evidence}\label{joint-evidence}

\begin{table*}[htbp]
\centering\small\setlength{\tabcolsep}{3pt}
\caption{Log evidence of the 20 scored DFT values (17 with the delta method) under each method\textquotesingle s
predictive distribution, in tolerance units (\(\varepsilon=10^{-6}\)), and the log Bayes factor against the headline with the same
covariance source. \(d^2\) and \(\log\det\) are the two parts of the evidence: the misfit of the errors, and the width of the covariance.}\label{tab:9}
\fitwidth{\begin{tabular}{@{}llllllll@{}}
\toprule
model & method & covariance from & \(n\) & \(\log p\) & log BF vs headline & \(d^2\) & \(\log\det\) \\
\midrule
o3d12 & CR3 × \(\lambda_F\) (headline) & committee & 20 & -45.0 & +0.0 & 49.6 & 3.6 \\
o3d12 & CR3 × \(\lambda_F\) (headline) & delta & 17 & -38.4 & +0.0 & 48.2 & -2.7 \\
o3d12 & B1: posterior × \(\hat q\) & committee & 20 & -59.2 & -14.2 & 80.1 & 1.6 \\
o3d12 & B1: posterior × \(\hat q\) & delta & 17 & -46.9 & -8.5 & 62.8 & -0.2 \\
o3d12 & B2: POPS (leverage cut 50 \%) & committee & 20 & -39.4 & +5.6 & 30.1 & 11.9 \\
o3d12 & B3: CR3, \(\lambda=1\) & committee & 20 & -827.5 & -782.5 & 1681.6 & -63.3 \\
o3d12 & B3: CR3, \(\lambda=1\) & delta & 17 & -803.9 & -765.6 & 1635.8 & -59.1 \\
o3d12 & CR3 × \(\lambda_V\) (sensitivity) & committee & 20 & -39.3 & +5.7 & 26.1 & 15.8 \\
o3d12 & CR3 × \(\lambda_V\) (sensitivity) & delta & 17 & -32.1 & +6.3 & 25.4 & 7.6 \\
o3d12 & CR3 × \(\lambda_F\) + discrepancy term & committee & 20 & -35.4 & +9.6 & 18.9 & 15.2 \\
o3d12 & CR3 × \(\lambda_F\) + discrepancy term & delta & 17 & -27.6 & +10.8 & 15.7 & 8.2 \\
o3d12 & B1 + discrepancy term & committee & 20 & -41.7 & +3.3 & 32.7 & 14.0 \\
o3d12 & B1 + discrepancy term & delta & 17 & -30.3 & +8.1 & 18.1 & 11.2 \\
o3d12 & B2 + discrepancy term & committee & 20 & -38.6 & +6.4 & 18.3 & 22.1 \\
o4d20 & CR3 × \(\lambda_F\) (headline) & committee & 20 & -11.3 & +0.0 & 22.8 & -37.0 \\
o4d20 & CR3 × \(\lambda_F\) (headline) & delta & 17 & -8.7 & +0.0 & 23.5 & -37.5 \\
o4d20 & B1: posterior × \(\hat q\) & committee & 20 & -40.2 & -29.0 & 100.7 & -56.9 \\
o4d20 & B1: posterior × \(\hat q\) & delta & 17 & -48.9 & -40.2 & 118.7 & -52.2 \\
o4d20 & B2: POPS (leverage cut 50 \%) & committee & 20 & -22.9 & -11.6 & 48.5 & -39.6 \\
o4d20 & B3: CR3, \(\lambda=1\) & committee & 20 & -398.3 & -387.0 & 865.9 & -106.1 \\
o4d20 & B3: CR3, \(\lambda=1\) & delta & 17 & -416.2 & -407.5 & 896.9 & -95.7 \\
o4d20 & CR3 × \(\lambda_V\) (sensitivity) & committee & 20 & -11.8 & -0.5 & 12.8 & -26.1 \\
o4d20 & CR3 × \(\lambda_V\) (sensitivity) & delta & 17 & -8.1 & +0.5 & 13.3 & -28.3 \\
o4d20 & CR3 × \(\lambda_F\) + discrepancy term & committee & 20 & -12.0 & -0.7 & 17.2 & -30.0 \\
o4d20 & CR3 × \(\lambda_F\) + discrepancy term & delta & 17 & -7.5 & +1.1 & 13.9 & -30.0 \\
o4d20 & B1 + discrepancy term & committee & 20 & -22.8 & -11.6 & 55.4 & -46.5 \\
o4d20 & B1 + discrepancy term & delta & 17 & -17.6 & -8.9 & 43.9 & -40.0 \\
o4d20 & B2 + discrepancy term & committee & 20 & -13.1 & -1.8 & 19.5 & -30.1 \\
\bottomrule
\end{tabular}}
\end{table*}

\begin{figure}[tbp]
\centering
\includegraphics[width=\columnwidth]{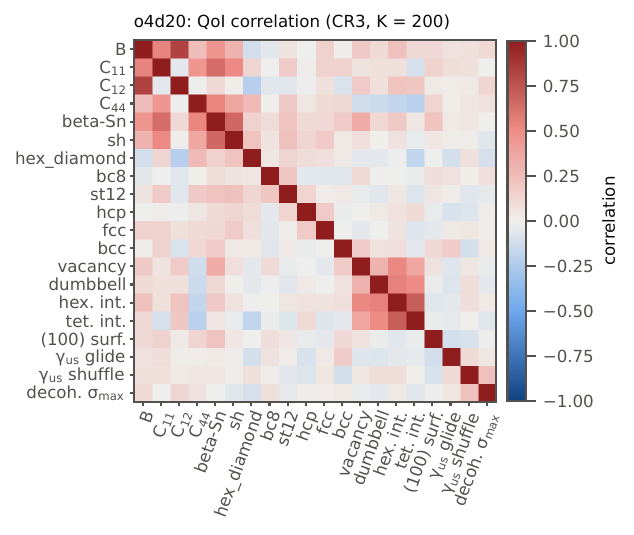}
\caption{Correlation
between the QoIs in the CR3 committee for o4d20 (\(K=200\) members).}\label{fig:7}
\end{figure}

The QoIs share one coefficient vector, so their errors are correlated, and Figure~\ref{fig:7} shows how. In the CR3 committee for o4d20 the
correlations form blocks: the elastic constants with each other and with the β-Sn and sh phase energies, which are both compressed
diamond-like structures; and the four point defects with each other (0.3--0.8), most strongly the three interstitials. Most other pairs
are only weakly correlated.

Table~\ref{tab:9} compares the methods by their evidence. The headline is far ahead of B1 and B3 for both models: by log Bayes
factors of -14.2 (B1) and -782.5 (B3) for o3d12, and -29.0 and
-387.0 for o4d20, with the committee covariance. On the scale of Kass and Raftery\cite{kass1995} a log Bayes factor
beyond about 3 is strong evidence and beyond 5 very strong. B3 loses through the misfit term, with intervals far too narrow for
its errors. B1 loses less, because its overall size is calibrated, but its shape puts the width in the wrong places. POPS (B2) is
the close competitor. Without the discrepancy term the evidence prefers it for o3d12 (+5.6) and the
headline for o4d20 (-11.6). The o3d12 preference does not survive the discrepancy term of
Section 3.8: with the term added to both, the headline is preferred for both models,
strongly for o3d12 (+3.1) and weakly for o4d20
(+1.1). One at a time the headline covers more QoIs for o3d12
(17/20 against 15/20) and POPS for o4d20
(17/20 against 15/20): per-QoI coverage and the joint
evidence need not agree. The correlations can also favour a method that
misses QoIs one at a time: o4d20\textquotesingle s headline misses five QoIs, but three of the five are the interstitials, which its covariance
expects to move together, so jointly they cost less than three independent misses.

The virial scale is better than the force scale for o3d12 (+5.7), whose QoI errors are larger
relative to its force errors, and about as good for o4d20 (-0.5). The discrepancy term of Section 3.8 is
strongly supported for o3d12 (+9.6 with the committee covariance,
+10.8 with the delta method), and roughly neutral for o4d20
(-0.7 and +1.1): o4d20\textquotesingle s defect bias is small, so the term\textquotesingle s
extra width costs about as much as the misses it removes. The delta-method and committee covariances agree, and the ranking of the
methods is the same for \(\varepsilon\) from \(10^{-8}\) to \(10^{-4}\).

\subsection{3.6 Delta method, committees and nonlinearity}\label{delta-method-committees-and-nonlinearity}

\begin{figure*}[tbp]
\centering
\includegraphics[width=\textwidth,height=0.6\textheight,keepaspectratio]{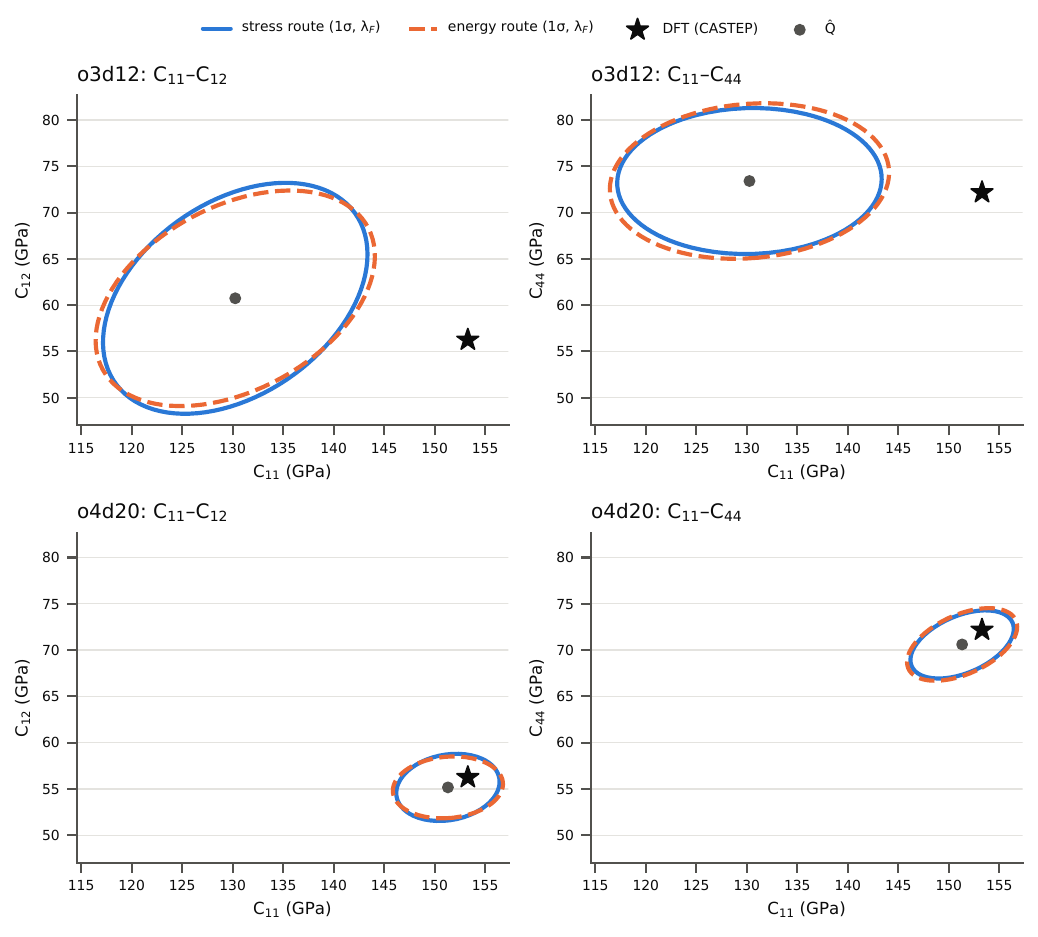}
\caption{Elastic constants: the \(1\sigma\) covariance ellipses of (\(C_{11}\), \(C_{12}\)) and (\(C_{11}\), \(C_{44}\)) at \(\lambda_F\) from
the delta method, by the stress route (solid) and the energy route (dashed), for o3d12 (top) and o4d20 (bottom); the dot is the
mean model\textquotesingle s value and the star DFT.}\label{fig:8}
\end{figure*}

The two routes to \(\sigma_Q\) agree where both apply. At \(\lambda=1\), the delta-method \(\sigma\) is within 10 \% of the committee\textquotesingle s for all but
1 of the 17 QoIs of o3d12 and 4 of the 17 of o4d20; the ratio ranges from
0.64 to 1.06 for o3d12 and from 0.79 to 1.16 for o4d20.
The flagged QoIs (the hcp phase energy for o3d12; the bulk modulus, the β-Sn phase energy, the vacancy, the tetrahedral interstitial for o4d20)
are taken from the committee, as Section 2.7.3 prescribes. The implicit differentiation through each relaxation, including that of the
lattice constant, is therefore accurate at the size of the CR3 distribution itself. The elastic constants are a stricter check,
because they can be computed by two routes, from the stress or from the curvature of the energy. Figure~\ref{fig:8} shows that the two give
nearly the same covariance ellipses for both models, and that the DFT values lie well outside the \(1\sigma\) ellipses for o3d12, whose
\(C_{11}\) is about 15 \% too low, but on or inside them for o4d20 (a \(1\sigma\) ellipse holds only 39 \% of a two-dimensional
Gaussian).

At \(\lambda_F\) the distribution is several times wider, and the linear rescaling of the \(\lambda=1\) committee,
\(Q_k(\lambda)=\hat Q+\lambda(Q_k-\hat Q)\), can fail. The full re-run at \(\lambda_F\) tests it. Of its 200 members,
180 (o3d12) and 198 (o4d20) pass the physicality rule fixed beforehand; the rest
collapse (20 and 2), and none fails a test. Outside the phase
energies, the \(\sigma\) of the physical members is 0.63--1.56× the rescaled \(\sigma\) for o3d12 and
0.80--1.49× for o4d20: for the elastic constants, defects and surfaces the rescaling is right to within these
factors. The phase energies are different: for o3d12 the \(\sigma\) of the β-Sn, sh and hcp energies is 37, 30 and 19 times the rescaled \(\sigma\), and for o4d20
the \(\sigma\) of the sh, β-Sn and hcp energies is 86, 43 and 24 times it. These are the phases furthest from diamond, whose structures relax into other phases when \(\theta\) moves far
enough, so the phase-energy \(\sigma\) at \(\lambda_F\) is a property of that relaxation path rather than of a Gaussian in \(\theta\). The headline
keeps the rescaled \(\sigma\), and Table~\ref{tab:8} reports the full re-run as a sensitivity.

\subsection{3.7 Baselines}\label{baselines-1}

B1 and B3 fall short in different ways (Figure~\ref{fig:5}, Figure~\ref{fig:6} and Table~\ref{tab:7}); B2 does not, once its box is numerically defined. B3, the CR3 shape without
calibration, covers only 8/20 (o3d12) and 5/20 (o4d20) of the QoIs at 95 \%: the shape alone is several times too narrow, as the test-split scales of
Section 3.3 already show. B1, the posterior with one conformal scale, covers 15/20 (o3d12) and 12/20 (o4d20). It has a calibrated overall size but the wrong
shape, so it is too wide for some QoIs and too narrow for others, and its joint evidence is far below the headline\textquotesingle s (Table~\ref{tab:9}).

B2, POPS, needed more care. A hypercube built from every training row, instead of the reference package\textquotesingle s 50 \% cut, gives
unphysical members: the bounds are set by a handful of rows with near-zero leverage, whose corrections are unbounded, and the members spread
the bulk energy of o3d12 by \(2\times10^{12}\) meV/atom. Rejecting unphysical members cannot rescue a box of that size.

The box is also numerically delicate on these designs. The o3d12 design matrix has a condition number of \(5\times10^{16}\), and
its evidence fit puts \(\sigma_c\) on its upper bound, so the ridge implied by the BLR prior is only \(4\times10^{-20}\) of the
largest eigenvalue. At ridges this small the box is not well defined: below a relative ridge of about \(10^{-11}\) it
depends on how it is computed. At \(10^{-14}\), mathematically equivalent evaluations from the same rows, mean and
leverage cut give o3d12 bulk-energy spreads of 1.6 (eigendecomposition), 2.1 (a singular value decomposition, SVD, of the rows) and 2.2 meV/atom (the hypercube step of the reference package POPSRegression\cite{popsregression}). We therefore select the ridge from \(10^{-10}\) upwards
(o3d12: E \(10^{-9}\), F \(10^{-10}\), V \(10^{-10}\); o4d20: E \(10^{-6}\), F \(10^{-8}\), V \(10^{-7}\)). The o3d12 selection lands on this floor or one decade above it, so for o3d12 the floor rather than the data largely
fixes the box; the o4d20 selection falls inside the grid. At the selected ridges the evaluations agree: the B2 box, the
eigendecomposition, the SVD and the reference package give covariances that agree to within \(7\times10^{-4}\) (o3d12) and \(2\times10^{-4}\) (o4d20), relative, so our implementation
and the reference package differ in numerics only, not in method. Two dependences remain. The size of the box depends strongly
on the ridge: for o3d12, with one ridge for all quantities, the bulk-energy spread is 4.54 meV/atom at \(10^{-10}\), 31 meV/atom at \(10^{-6}\), 263 meV/atom at \(10^{-2}\). It also depends on the
coordinates: the reference package takes its axes in the coordinates it is given, here the prior-scaled \(\Gamma\theta\), and its box there
differs from the box in \(\theta\) (161 against 206 modes, 4.53 against 5.50 meV/atom (o3d12) and 683 against 2766 modes, 0.87 against 1.10 meV/atom (o4d20) in \(\theta\) against \(\Gamma\theta\), at the selected ridges). The reference package\textquotesingle s default ridge, chosen by evidence, is \(4\times10^{-21}\) (o3d12) of the
largest eigenvalue, below the floor, so we do not use it.

We ran committees at the 50 \% cut (B2) and at 5 \%, at the ridges selected per quantity; Table~\ref{tab:10} compares them with the
CR3 members. The 5 \% cut widens the box, by a factor of 1.4 (o3d12) and 2.4 (o4d20) in bulk-energy spread, but it changes the 95 \% coverage
by at most one QoI. With a numerically defined box, POPS gives QoI intervals of about the headline\textquotesingle s width and
coverage without a calibration step. What sets its width is the ridge and the leverage cut instead: over the numerically defined
ridges the o3d12 box\textquotesingle s bulk-energy spread changes by a factor of about 60 (above).

\begin{table*}[htbp]
\centering\small\setlength{\tabcolsep}{3pt}
\caption{POPS committees at the two leverage cuts against the CR3 members: spread of the
members\textquotesingle{} bulk (diamond) energy, and QoI coverage and sharpness as in Table~\ref{tab:7}.}\label{tab:10}
\fitwidth{\begin{tabular}{@{}lllllll@{}}
\toprule
model & method & bulk-energy spread (meV/atom) & 95 \% & RMS \(z\) & median \(\sigma/\text{tol}\) & CRPS/tol \\
\midrule
o3d12 & B2: POPS, leverage cut 50 \% & 3.89 & 15/20 & 1.55 & 1.75 & 2.86 \\
o3d12 & POPS, leverage cut 5 \% & 5.64 & 15/20 & 1.42 & 2.44 & 2.87 \\
o3d12 & CR3 × \(\lambda_F\) (headline) & 1.98 & 17/20 & 1.29 & 2.52 & 2.79 \\
o3d12 & B3: CR3, \(\lambda=1\) & 0.34 & 8/20 & 7.53 & 0.43 & 3.16 \\
o4d20 & B2: POPS, leverage cut 50 \% & 0.83 & 17/20 & 1.69 & 0.60 & 0.81 \\
o4d20 & POPS, leverage cut 5 \% & 2.02 & 18/20 & 1.11 & 0.99 & 0.84 \\
o4d20 & CR3 × \(\lambda_F\) (headline) & 0.69 & 15/20 & 1.31 & 0.64 & 0.75 \\
o4d20 & B3: CR3, \(\lambda=1\) & 0.11 & 5/20 & 8.11 & 0.10 & 0.91 \\
\bottomrule
\end{tabular}}
\end{table*}

\protect\phantomsection\label{results-defects}
\subsection{3.8 Defect formation energies: diagnosis and the discrepancy term}\label{defect-formation-energies-diagnosis-and-the-discrepancy-term}

For both models the relaxed interstitial formation energies are the QoIs that are confidently wrong: all three are too
low, and outside the 95 \% interval. This holds for every method of Section 3.7, not only the headline, so the diagnosis below
is about the fit, not the uncertainty method; the discrepancy term it leads to (below) applies to any of them. Table~\ref{tab:13} shows it for
the posterior baseline B1 as well as for CR3: with the term B1 covers 18/20 (o3d12) and 16/20 (o4d20). They are low for every ACE model in the benchmark too, linear and nonlinear (PACE), and the tetrahedral one
for GAP-18.

\begin{table}[htbp]
\centering\small\setlength{\tabcolsep}{3pt}
\caption{The training residual trend predicts the QoI error. Interstitial cells,
residual (DFT − model) minus a temperature-matched bulk reference, extrapolated to \(F_{\text{rms}}\to0\) (eV), against the QoI
errors (DFT − model) of the dumbbell / hexagonal / tetrahedral interstitials, and the mean over the hot training cells.}\label{tab:11}
\fitwidth{\begin{tabular}{@{}llll@{}}
\toprule
model & extrapolated to \(F_{\text{rms}}\to0\) & QoI errors & mean over hot cells \\
\midrule
o3d12 & +1.58 ± 0.06 & +1.18 / +1.45 / +1.74 & -0.46 \\
o4d20 & +0.38 ± 0.03 & +0.37 / +0.41 / +0.60 & -0.43 \\
\bottomrule
\end{tabular}}
\end{table}

Over the hot cells the model\textquotesingle s interstitial error averages near zero, so the CR3 intervals are calibrated there (test-split
\(z\) RMS 0.85--1.13 for those cells). The error grows towards the cold limit, and the relaxed defect lies beyond the data along
a bias that all interstitial clusters share. With per-atom energy weighting, 0.4--1.6 eV in a 217-atom cell costs only
2--7 meV/atom in the fit.

\textbf{Why not in parameter space.} The interstitial synthetic row is far outside the jackknife\textquotesingle s reach: its CR3 × \(\lambda_F\) \(\sigma\) is
0.49 eV against \(r^*=1.58\) eV (o3d12) and 0.17 eV against \(r^*=0.38\) eV (o4d20). The minimum-norm correction that removes \(r^*_t\), \(\delta_t=\Sigma d^*_t r^*_t/(d^{*\mathsf T}_t\Sigma d^*_t)\), with \(\Sigma\) the
propagated covariance (the POPS construction applied to the synthetic row, in the CR3 or posterior metric), added as
\(\Sigma\mathrel{+}=\delta_t\delta_t^{\mathsf T}\), carries part of the bias to the relaxed defects, but unevenly: the projected \(\sigma\) on the three
interstitial QoIs is o3d12: 0.2--1.4 of \(r^*=1.58\) eV; o4d20: 0.3--0.7 of \(r^*=0.38\) eV (both metrics). The hot-snapshot contrast and the relaxed defect\textquotesingle s formation row point
in different directions in \(\theta\), so the transfer depends on the QoI. Coverage at 95 \% with \(\lambda_F\), over the 17 QoIs the delta method gives, becomes o3d12 14 → 14--16 of 17, o4d20 12 → 15--16 of 17,
short of the observable-space term below.

\protect\phantomsection\label{results-defects2}
\begin{table}[htbp]
\centering\small\setlength{\tabcolsep}{3pt}
\caption{The bias \(r^*_t\) and its standard error s*\textsubscript{t} in the headline discrepancy term
(Section 2.7.4; DFT training labels only). Every row is an extrapolation of the training residuals to zero thermal amplitude
(η → 0). For the vacancy that extrapolation is also for the size of the QoI cell (215 atoms), averaged over the four size
models of Table~\ref{tab:15}, and s* is the standard deviation of that average.}\label{tab:12}
\fitwidth{\begin{tabular}{@{}lllll@{}}
\toprule
model & class & estimator & \(r^*\) (eV) & s* (eV) \\
\midrule
o3d12 & interstitial & bulk-trend regression, η → 0 & +1.577 & 0.062 \\
o3d12 & vacancy & 215-atom intercept, η → 0, averaged over size models & +0.214 & 0.137 \\
o4d20 & interstitial & bulk-trend regression, η → 0 & +0.385 & 0.032 \\
o4d20 & vacancy & 215-atom intercept, η → 0, averaged over size models & -0.042 & 0.049 \\
\bottomrule
\end{tabular}}
\end{table}

\begin{table*}[htbp]
\centering\small\setlength{\tabcolsep}{3pt}
\caption{Coverage of the 20 scored QoIs with the discrepancy term on the four defect QoIs
(models with \(r^*\): o3d12, o4d20).}\label{tab:13}
\fitwidth{\begin{tabular}{@{}lllllll@{}}
\toprule
model & method & 95 \% without & 95 \% with {[}credible interval{]} & 68 \% & median \(\sigma/\text{tol}\) & CRPS/tol without → with \\
\midrule
o3d12 & CR3 × \(\lambda_F\) & 17/20 & 20/20 {[}0.88, 1.00{]} & 14 → 17 & 2.52 & 2.79 → 2.38 \\
o3d12 & CR3 × \(\lambda_V\) & 18/20 & 20/20 {[}0.88, 1.00{]} & 15 → 18 & 3.47 & 2.78 → 2.53 \\
o3d12 & B1: posterior × \(\hat q\) & 15/20 & 18/20 {[}0.72, 0.98{]} & 13 → 15 & 1.66 & 2.92 → 2.30 \\
o4d20 & CR3 × \(\lambda_F\) & 15/20 & 18/20 {[}0.72, 0.98{]} & 12 → 14 & 0.64 & 0.75 → 0.65 \\
o4d20 & CR3 × \(\lambda_V\) & 19/20 & 20/20 {[}0.88, 1.00{]} & 14 → 16 & 0.85 & 0.73 → 0.66 \\
o4d20 & B1: posterior × \(\hat q\) & 12/20 & 16/20 {[}0.59, 0.93{]} & 11 → 12 & 0.39 & 0.82 → 0.64 \\
\bottomrule
\end{tabular}}
\end{table*}

The term lowers the CRPS of every method (Table~\ref{tab:13}), most for o3d12, whose interstitial errors dominate its score. With the
term, B1 and CR3 × \(\lambda_F\) score almost the same (2.30 against
2.38 for o3d12, 0.64 against
0.65 for o4d20, B1 marginally lower). Once the defects carry the term, the remaining
difference between the two shapes is small on this score, while CR3 still covers more QoIs.

The interstitial estimator agrees well with the observed interstitial errors. That agreement was seen before the bulk
reference was chosen: an alternative, each defect cell\textquotesingle s nearest bulk cell in volume and thermal amplitude, was rejected
because its standard error is larger and it extrapolates worse to held-out cold snapshots. The choice was therefore not made
blind, although the reasons for it use training data only.

\begin{table*}[htbp]
\centering\small\setlength{\tabcolsep}{3pt}
\caption{Defect QoIs, CR3 × \(\lambda_F\): error, error of the prediction corrected by \(r^*\) (for
reference; not used), \(\sigma\) without and with the discrepancy term (eV), and \(z\).}\label{tab:14}
\fitwidth{\begin{tabular}{@{}lllllll@{}}
\toprule
model & QoI & DFT − model & DFT − (model + \(r^*\)) & \(\sigma\) & \(\sigma\) with term & \(z\) \\
\midrule
o3d12 & vacancy & +0.881 & +0.667 & 0.545 & 0.601 & -1.62 → -1.47 \\
o3d12 & dumbbell int. & +1.181 & -0.396 & 0.351 & 1.617 & -3.37 → -0.73 \\
o3d12 & hexagonal int. & +1.448 & -0.129 & 0.525 & 1.663 & -2.76 → -0.87 \\
o3d12 & tetrahedral int. & +1.738 & +0.161 & 0.847 & 1.791 & -2.05 → -0.97 \\
o4d20 & vacancy & +0.111 & +0.154 & 0.084 & 0.106 & -1.33 → -1.05 \\
o4d20 & dumbbell int. & +0.372 & -0.013 & 0.177 & 0.425 & -2.10 → -0.88 \\
o4d20 & hexagonal int. & +0.412 & +0.028 & 0.176 & 0.424 & -2.34 → -0.97 \\
o4d20 & tetrahedral int. & +0.599 & +0.214 & 0.220 & 0.444 & -2.72 → -1.35 \\
\bottomrule
\end{tabular}}
\end{table*}

\protect\phantomsection\label{results-vacancy}
\subsubsection{The vacancy}\label{the-vacancy}

The vacancy shows no thermal trend when its 63- and 215-atom training cells are pooled. Table~\ref{tab:15} compares the size models of the
method above by their evidence, from training labels only.

\begin{table*}[htbp]
\centering\small\setlength{\tabcolsep}{3pt}
\caption{Vacancy size models: log Bayes factor against the best model, and the posterior
zero-thermal-amplitude bias of a 215-atom cell (eV), for each model and averaged over the models (\(s_a=1\) eV). The
last pair of columns applies the same analysis to the energy difference of the two models on the training cells (o4d20 − o3d12);
the true difference of the models\textquotesingle{} own relaxed vacancy QoIs is 0.77 eV (no DFT value involved).}\label{tab:15}
\fitwidth{\begin{tabular}{@{}lllllll@{}}
\toprule
model of the size dependence & o3d12 log BF & bias & o4d20 log BF & bias & model vs model log BF & bias \\
\midrule
one intercept, one slope & -114.6 & +0.79 ± 0.04 & -142.7 & +0.14 ± 0.02 & -80.1 & +0.60 ± 0.04 \\
per-size intercepts, one slope & -3.9 & -0.20 ± 0.04 & -15.1 & -0.33 ± 0.02 & -1.0 & +0.14 ± 0.03 \\
one intercept, per-size slopes & -9.0 & +0.79 ± 0.04 & -5.4 & +0.14 ± 0.02 & -2.6 & +0.64 ± 0.03 \\
per-size intercepts and slopes & +0.0 & +0.22 ± 0.12 & +0.0 & -0.04 ± 0.05 & +0.0 & +0.35 ± 0.10 \\
average over models & & +0.21 ± 0.14 & & -0.04 ± 0.05 & & +0.31 ± 0.14 \\
\bottomrule
\end{tabular}}
\end{table*}

The data reject one trend for both sizes outright, and disfavour a shared intercept: the zero-temperature bias depends on the
cell size. The per-size models win, and the vacancy QoI, in a 216-site cell, takes the 215-atom intercept averaged over the
models: +0.21 ± 0.14 eV (o3d12) and
-0.04 ± 0.05 eV (o4d20). The 215-atom cells are all hot, so
this extrapolation is weakly determined and depends on the prior. For the shared-intercept model, the log Bayes factor and the model average at the other prior
scales are, for o3d12, \(s_a=0.3\) eV: -74.4, average -0.23 eV; \(s_a=3\) eV: -3.6, average +0.35 eV; for o4d20, \(s_a=0.3\) eV: -16.1, average -0.12 eV; \(s_a=3\) eV: -3.5, average -0.03 eV.

One check disagrees. Applied to the difference between the two models, where the truth is known without DFT, the
shared-intercept model with per-size slopes predicts 0.64 of the true 0.77 eV and the model average only
0.31. Both miss by several standard errors, so neither
form captures how the relaxed vacancy differs from the hot cells. The one near-relaxed training cell (63 atoms) gives
direct evidence without extrapolation: its bulk-subtracted residual is +0.61 eV (o3d12) and 0.00 eV (o4d20), and relaxing it
(o3d12: relaxation gain 0.34 eV against a DFT bound of 0.24 eV, o4d20: relaxation gain 0.18 eV against a DFT bound of 0.18 eV) suggests that o3d12 over-relaxes. Both models relax the harness vacancy to a symmetric structure (vacancy neighbour
distances o3d12 3.40--3.40 Å, o4d20 3.47--3.47 Å); the harness gives no DFT structure to compare.

The vacancy term is small in both models, and the CR3 \(\sigma\) alone already covers the vacancy at 95 \% (Table~\ref{tab:14}).

\begin{figure*}[tbp]
\centering
\includegraphics[width=\textwidth,height=0.6\textheight,keepaspectratio]{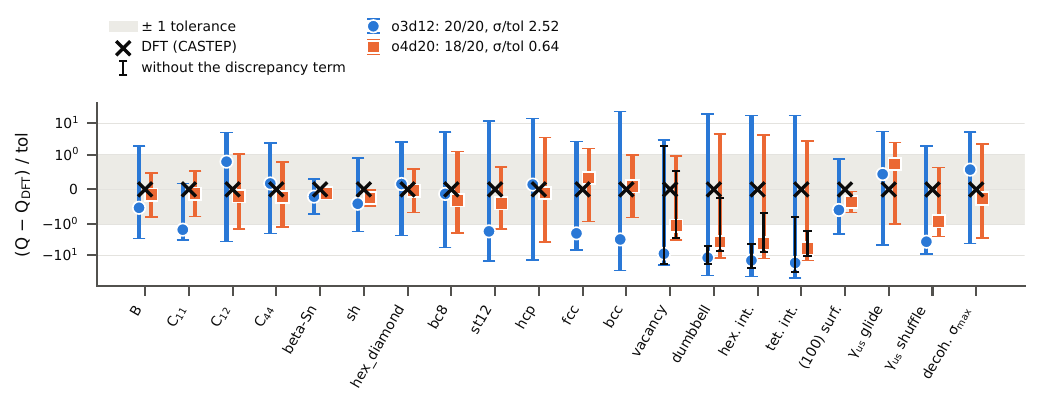}
\caption{As Figure~\ref{fig:5}, with the headline discrepancy term on the four defect QoIs (CR3 × \(\lambda_F\)): the bulk-trend regression for the
interstitials and the size-model average for the vacancy. Thin black bars: the same QoIs\textquotesingle{} 95 \% intervals without the term.
Only the four defect QoIs change.}\label{fig:9}
\end{figure*}

\section{4 Discussion}\label{discussion}

A single parameter covariance, the CR3 jackknife scaled once to held-out forces, gives usable uncertainties for derived
properties that the fit never saw. For both models most QoIs fall inside their 95 \% intervals (17/20
for o3d12, 15/20 for o4d20), and the intervals scale with the model: the median \(\sigma/\text{tol}\) is
2.52 for o3d12 and 0.64 for the accurate o4d20, whose
intervals are mostly inside the benchmark tolerance. The scale is fitted once, on forces of held-out configurations, and never on a QoI.
The same CR3 shape serves per-atom forces, with one conformal quantile (Section 3.2), and relaxed QoIs, with one global
scale. The second use is what a distribution over \(\theta\) buys: one draw gives energies, forces, stresses and every QoI together,
consistently.

The baselines show which ingredient matters. The CR3 shape without a scale (B3) is far too narrow, so calibration is essential,
as expected for a parameter variance that shrinks with the amount of data. The Bayesian posterior with a calibrated scale (B1)
has the right overall size but the wrong shape: it under-covers the QoIs and its joint evidence is far lower, because the posterior describes
noise in the labels, not the approximation error of a misspecified model. Shanks et al.\cite{shanks2026} find the same split between
accuracy and uncertainty for argon: MACE committees and seed ensembles predict accurately but cover far fewer errors than their
intervals claim, and calibration, sharpness and the ranking of errors need not improve together. POPS (B2) targets that error directly and, with a numerically defined box,
performs comparably on these QoIs: similar coverage and widths. The joint evidence is mixed: without the discrepancy term it
prefers POPS for o3d12 (log Bayes factor +5.6) and the CR3 propagation for o4d20
(+11.6); with the term on both it prefers the CR3 propagation, by
+3.1 (o3d12) and +1.1 (o4d20). The CR3 approach is therefore not more accurate than POPS.
Implicit differentiation is no distinction either: Perez et al.\cite{perez2025} combine it with POPS, and either distribution can be
propagated by the delta method or by committees. What CR3 adds is a parameter distribution that is calibrated and numerically
well posed.

Three properties of the distribution make the difference. First, its size is one scale, fitted to held-out errors and
interpretable as such. The size of the POPS box is set instead by the ridge and the leverage cut: across the numerically defined
ridges the o3d12 box\textquotesingle s bulk-energy spread changes by a factor of about 60, the held-out selection put the o3d12 ridge at the floor,
and a box built from every row is unphysical. Second, a covariance is a sum over configurations, and it is formed here through its
factor \(L\), never as an ill-conditioned quadratic form. A hypercube is bounded by the extreme projections of single rows: on the same
rows, mean and cut, equivalent evaluations of one box gave bulk-energy spreads from 1.6 to 2.2 meV/atom at a relative ridge of \(10^{-14}\), and the box exists only above a ridge floor. Third, a
covariance transforms consistently under a linear change of parameters, and a box does not: with the axes taken in \(\theta\) or in the
prior-scaled coefficients \(\Gamma\theta\), the same corrections give different boxes (161 against 206 modes, 4.53 against 5.50 meV/atom (o3d12) and 683 against 2766 modes, 0.87 against 1.10 meV/atom (o4d20) in \(\theta\) against \(\Gamma\theta\), at the selected ridges). For forces the calibration is explicit. One conformal
quantile on the CR3 shape covers 0.87 and 0.90 of the test atoms against a nominal 0.90 (the
o3d12 shortfall is the hold-out transfer of Section 3.2), and holds out of distribution on the cracks for both models and on
the screw dislocations for o4d20, though not for o3d12 (0.77); the global force scale on the same
kind of region covers less (0.79 and 0.88), even fitted in sample. The POPS box covers
0.95 (o3d12) and 0.84 (o4d20), the latter below nominal and unevenly (per type 0.51 (surface\_001) to 0.99), and nothing in its construction aims at a target coverage.
Refitted without the cracks and dislocations, it covers 0.89 of the held-out atoms for o3d12 but 0.59 for o4d20, against
0.76 and 0.91 for CR3 × \(\lambda_F\) (Section 3.4). Two costs of CR3 remain. It is a sampling variance, so it needs held-out data and a
scale; POPS needs neither in principle, although here its ridge was chosen on a hold-out of the training set. And like POPS it cannot see a bias that every configuration of a kind shares (Section 3.8).

The scale is not unique. The virial scale is about a third larger than the force scale, and using it covers
18/20 QoIs for o3d12 and 19/20 for o4d20 with wider intervals; its joint evidence is higher than the force scale\textquotesingle s for o3d12 and about equal for o4d20. Misspecification depends on the observable, and one could choose the scale by the QoI\textquotesingle s physics, the virial scale
for stress-like QoIs such as elastic constants, or report both. A single scale is also a compromise across configuration types
(Figure~\ref{fig:5}), which a per-type scale in \(\theta\) could relax. A learned per-atom scale cannot be used for this directly: it rescales
force outputs atom by atom and defines no covariance in \(\theta\); a \(\theta\)-space analogue would need a shape that depends on the
environment.

Linearisation is adequate at the size of the distribution, but not everywhere at the calibrated size. The delta method agrees
with committees run through the unmodified benchmark for almost every QoI at \(\lambda=1\), and implicit differentiation through each
relaxation, including that of the lattice constant, makes it cheap: one set of rows per QoI. At \(\lambda_F\) the phase energies of
the structures furthest from diamond become strongly nonlinear, because their relaxations change character; for those, committees
are the safer route, and their spread is then a property of the relaxation path rather than of a Gaussian.

The clearest failure is the relaxed point defects, missed by every method. Its cause is in the data: every defect cell in
GAP-18 is a hot molecular-dynamics snapshot, the model\textquotesingle s defect--bulk error drifts with thermal amplitude, and the relaxed defect
lies beyond the data along a direction that every snapshot shares. A delete-one jackknife cannot see such a bias, because deleting
one snapshot does not move the fit along it, and a rank-one parameter correction built from the snapshots carries it to the relaxed
geometry only partly and unevenly. The bias can, however, be measured from the training labels alone, by extrapolating the defect--bulk
residual to zero thermal amplitude, and added as a discrepancy term; this raises the coverage to 20/20 (o3d12) and
18/20 (o4d20) without using any DFT QoI value, although the term was devised after the QoI misses had been seen. For the vacancy, Bayesian comparison of models for the training cells\textquotesingle{}
size dependence shows that the bias depends on the cell size; the 215-atom extrapolation it selects is weakly determined,
a reminder that such extrapolations depend on how the training data were made. Relaxed or low-temperature defect
configurations in the training set would remove the need for the term, and the same diagnostic, a residual trend that extrapolates
away from zero, can flag where such data are missing.

Mondrian conformal prediction\cite{vovk2005}, which calibrates a separate quantile in each group of a
partition fixed in advance, is a natural extension for environments absent from training. The \texttt{ace-jax} force uncertainty
provides eight such groups of atoms, set by the distortion and coordination of each atom\textquotesingle s first shell\cite{acejax}. On a
Cantor-alloy model in the package\textquotesingle s development tests\cite{acejax_cantor}, those groups with the hold-out transfer met their coverage targets on crack and dislocation cells left out of both training
and calibration: 0.90 on the cracks, 0.89 at the crack tips and 0.94 on edge and screw dislocations, at a nominal 0.90.
That test did not compare the groups with a single quantile. On silicon here the same groups change the out-of-distribution
coverage of the atom-weighted quantile by at most 0.02 and leave the screw dislocations of o3d12 under-covered
(0.79, against 0.77 with one quantile).
Whether groups resolve the remaining under-coverage on strongly distorted environments, for example with finer distortion
bands or a descriptor-novelty axis, is left to future work.

The study has limitations. The calibration assumes that the test split and the QoI structures are exchangeable with the
training data, which the defects show is not always true. The hyperparameters are held fixed in the jackknife, so their own
uncertainty is not propagated. The locality of the model, which cannot represent interactions beyond its cutoff, appears here only
as part of the effective error. And twenty QoIs give coarse coverage estimates: 17 of 20 leaves a 95 \% credible interval from
0.65 to 0.96 for the coverage probability, so the evidence of Section 3.5, which uses all the QoIs jointly, is the sharper comparison.

\section{5 Conclusions}\label{conclusions}

For linear ACE potentials, the delete-one-configuration (CR3) jackknife covariance, scaled once, gives uncertainties close
to their target coverage at two levels. Per atom, with one conformal quantile from a held-out part of the training set, the
force regions come close to their target coverage on held-out silicon data and on cracks left out of the fit, though for the
smaller model screw dislocations left out are under-covered; a scale learned from the site descriptor evens the coverage across
environments seen in training but does not improve it on those left out, and for the larger model lowers it. For derived properties, the same covariance with one scale fitted to held-out forces,
propagated by implicit differentiation through each relaxation, covers most of 20 silicon QoIs against DFT, with intervals that
shrink with the accuracy of the model. It outperforms the Bayesian posterior and the uncalibrated shape, and on these QoIs it performs
comparably with POPS once the POPS box is numerically defined. Its advantage over POPS is not accuracy but calibration and numerical
robustness: one interpretable scale, fitted to held-out errors, sets the width at both levels; the covariance needs no ridge or leverage
cut beyond the fit\textquotesingle s own prior, and does not depend on the parameterisation; and the per-atom regions are calibrated to a target
coverage. On per-atom forces left out of the fit, POPS is not consistent between the two models: refitted without the cracks
and dislocations, it covers 0.89 of the held-out atoms for o3d12, above the 0.76 of the scaled CR3 covariance, but
0.59 for o4d20, where the scaled CR3 covariance covers 0.91. Where the CR3 covariance fails, for
relaxed point defects, the cause is a bias shared by all training snapshots of a kind, which no delete-one method can see; it can be
measured from the training labels and added as a discrepancy term, which brings the coverage close to nominal. The methods are
implemented in \texttt{ace-jax}.

\protect\phantomsection\label{methods-detail}
\section{6 Computational details}\label{computational-details}

The training and test data are the GAP-18 silicon database\cite{bartok2018}, split once at random into 2227 training and
248 test configurations (seed 0; the split\textquotesingle s content hash is c433c98d49bb). Both models are linear ACE potentials fitted with
\texttt{ace-jax}\cite{acejax} at a 5 Å cutoff, with least-squares reference energies, the smoothness prior of Section 2.2, one shared noise scale
and the structural weights of the \texttt{ace-jax} silicon accuracy--cost benchmark. Every calculation is in double precision. The QoIs are
computed by a Python 3 port of the silicon test framework of Bartók et al.\cite{bartok2018}, run unmodified apart from a relaxed-ion force
tolerance of \(10^{-3}\) eV/Å for the elastic constants. Committees use 200 members drawn with fixed seeds. The fits, committees and analyses ran on
workstations of the University of Warwick\textquotesingle s Scientific Computing Research Technology Platform, and the cost timings of Figure~\ref{fig:1} on
a CPU node and an NVIDIA A100 GPU of the Sulis Tier 2 HPC platform.

\textbf{What was fixed in advance.} The specification of the study, committed before any QoI was computed (repository below),
fixed the 21 QoIs and their tolerances, the four candidate scales \(\lambda_F\), \(\lambda_E\), \(\lambda_V\) and
\(\lambda_{\text{pool}}\), the three baselines, the committee size and the 10 \% rule that picks the delta method or the committee. \(\lambda_F\)
was chosen as the headline before any QoI coverage was seen; the specification lists the four scales but does not record that
choice. Decided later, after QoI results had been seen: the POPS leverage cut (the box was first built from every row, which gave unphysical
members, Section 3.7; B2 follows the reference package\textquotesingle s default, the 50 \% cut) and its ridge floor (set by where the box is numerically
defined, Section 3.7), the member-rejection rule of the \(\lambda_F\) re-run (fixed
before that re-run), the discrepancy term and its estimators (Section 2.7.4; computed from the training cells only, the test split\textquotesingle s defect
cells being left out), the per-atom force method (Section 2.5, chosen after
comparing it with a learned scale on the test split and on the out-of-distribution set), the use of the BLR posterior instead
of ARD for that stage (so that one posterior serves forces and QoIs; Section 2.3 gives both), and Bayesian model comparison in place of
the specification\textquotesingle s fixed {[}0.8, 1.25{]} band on \(\lambda_V/\lambda_F\). None of these uses a DFT QoI value as an input, but
they were not made blind.

\textbf{Use of AI tools in the research.} The analysis software of this study was written and run with Claude Code (Anthropic;
model Claude Opus 5.5): the uncertainty propagation, calibration and validation code in \texttt{ace-jax} and the scripts that produce
every table and figure. Claude Code was used to implement the analysis the author specified, to carry out the runs and to draft
text. It did not generate or alter any data: every number comes from the stored results of that code. The author checked the
code, the results and the text, including the parity tests and finite-difference checks of Section 2.7.2. Session transcripts
are available on request.

\textbf{Code and data availability.} The uncertainty methods, including the per-atom stage at fixed hyperparameters
(\texttt{-\/-ard-prior\ fixed}), are part of \texttt{ace-jax} version 0.5.3 (MIT licence, \url{https://github.com/ACEsuit/ace-jax}); the
per-atom fits were run from its development branch before the release, and the parity and transfer analyses of Section 3.2
reproduce exactly with the release. The analysis code, the results behind every table and figure, the generator of this paper and
the original specification of the study are at \url{https://github.com/kermodegroup/si-qoi-uq-paper}; the version used here, release v1.0.0, is archived on
Zenodo at \url{https://doi.org/10.5281/zenodo.23986930}. The GAP-18 database is available from its original publication\cite{bartok2018}. The Python 3 port of the silicon test
framework used here is the branch py3 of \url{https://github.com/libAtoms/silicon-testing-framework}, at commit f23b1df.

\protect\phantomsection\label{ack}
\section{Acknowledgements}\label{acknowledgements}

The author thanks Thomas Swinburne and George Simmons for useful discussions.

Calculations were performed on the Sulis Tier 2 HPC platform hosted by the Scientific Computing Research
Technology Platform at the University of Warwick; Sulis is funded by EPSRC Grant EP/T022108/1 and the HPC Midlands+ consortium. We also used the Scientific Computing Research Technology Platform\textquotesingle s desktop compute nodes.

\textbf{Use of AI tools.} Claude Code (Anthropic; model Claude Opus 5.5) was used to write and run the analysis software in this work (the propagation,
calibration and validation code in the \texttt{ace-jax} repository and the scripts that produce every table and figure), and to help draft
and revise the text. All numbers in the paper are generated by that code from the stored results. Claude Code usage for software
development was supported by credits from Anthropic\textquotesingle s AI for Science program. The author designed the study, reviewed the code,
results and text, and takes full responsibility for the correctness of the resulting implementation and for the content of this
paper.

\bibliography{references}
\end{document}